\documentclass[twocolumn]{aastex63}
\usepackage{amsmath,bm}
\usepackage{xcolor}
\usepackage[english]{babel}
\usepackage[latin1]{inputenc} 
\usepackage{enumitem}

\newcommand{\kms}{$\mathrm{km\,s^{-1}}$}
\newcommand{\msun}{M_{\sun}}
\newcommand{\msunh}{h^{-1}M_{\sun}}

\newcommand{\hi}{H{\sc i}}
\newcommand{\hj}{$\rm{H_2}$}

\shorttitle{Central Galaxy Star Formation and Quenching}
\shortauthors{H. Guo et al.}

\begin{document}

\title{Star Formation and Quenching of Central Galaxies from Stacked \hi\ Measurements}

\author[0000-0003-4936-8247]{Hong Guo}
\affiliation{Key Laboratory for Research in Galaxies and Cosmology, Shanghai Astronomical Observatory, Shanghai 200030, China}

\author[0000-0002-5434-4904]{Michael G. Jones}
\affiliation{Steward Observatory, University of Arizona, 933 North Cherry Avenue, Rm. N204, Tucson, AZ 85721-0065, USA}
\affiliation{Instituto de Astrof\'isica de Andaluc\'ia, Glorieta de la Astronom\'ia s/n, 18008 Granada, Spain}

\author[0000-0002-6593-8820]{Jing Wang}
\affiliation{Kavli Institute for Astronomy and Astrophysics, Peking University, Beijing 100871, China}

\author[0000-0003-1138-8146]{Lin Lin}
\affiliation{Key Laboratory for Research in Galaxies and Cosmology, Shanghai Astronomical Observatory, Shanghai 200030, China}

\correspondingauthor{Hong Guo, Michael G. Jones, Jing Wang}
\email{guohong@shao.ac.cn;\\
	jonesmg@arizona.edu;\\
	jwang\_astro@pku.edu.cn}

\begin{abstract}
	We quantitatively investigate the dependence of central galaxy \hi\ mass ($M_{\rm HI}$) on the stellar mass ($M_\ast$), halo mass ($M_{\rm h}$), star formation rate (SFR), and central stellar surface density within 1~kpc ($\Sigma_1$), taking advantage of the \hi\ spectra stacking technique using both the Arecibo Fast Legacy ALFA Survey and the Sloan Digital Sky Survey. We find that the shapes of $M_{\rm HI}$--$M_{\rm h}$ and $M_{\rm HI}$--$M_\ast$ relations are remarkably similar for both star-forming and quenched galaxies, with massive quenched galaxies having constantly lower \hi\ masses of around 0.6~dex. This similarity strongly suggests that neither halo mass nor stellar mass is the direct cause of quenching, but rather the depletion of \hi\ reservoir. While the \hi\ reservoir for low-mass galaxies of $M_\ast<10^{10.5}\msun$ strongly increases with $M_{\rm h}$, more massive galaxies show no significant dependence of $M_{\rm HI}$ with $M_{\rm h}$, indicating the effect of halo to determine the smooth cold gas accretion. We find that the star formation and quenching of central galaxies are directly regulated by the available \hi\ reservoir, with an average relation of ${\rm SFR}\propto M_{\rm HI}^{2.75}/M_\ast^{0.40}$, implying a quasi-steady state of star formation. We further confirm that galaxies are depleted of their \hi\ reservoir once they drop off the star-formation main sequence and there is a very tight and consistent correlation between $M_{\rm HI}$ and $\Sigma_1$ in this phase, with $M_{\rm HI}\propto\Sigma_1^{-2}$. This result is in consistent with the compaction-triggered quenching scenario, with galaxies going through three evolutionary phases of cold gas accretion, compaction and post-compaction, and quenching.  
\end{abstract}

\section{Introduction}
Galaxies in the local Universe are known to consist of two broad categories of early and late types and their  color and SFR distributions are found to be bimodal \citep[see e.g.,][]{Baldry2004,Balogh2004,Brinchmann2004}. Galaxies are then naturally separated into the red/blue or the star-forming/quenched populations. One of the fundamental questions in modern extragalactic astronomy is how galaxies evolve from the bluer and younger population to that of the redder and older one, i.e. how galaxies quench their star formation.  

Various physical mechanisms have been proposed to understand this process, such as the virial shock heating \citep[e.g.,][]{Rees1977, Silk1977, White1978, Birnboim2003, Keres2005}, stellar feedback \citep[e.g.,][]{Dekel1986,Murray2005}, active galactic nuclei (AGN) feedback \citep[e.g.,][and references therein]{Fabian2012}, violent disk instabilities \citep[e.g.][]{Gammie2001,Dekel2009,Cacciato2012}, and morphological quenching \citep{Martig2009}. As discussed in the review of \cite{Man2018}, most of these mechanisms are trying to explain the cooling, inflow and outflow of the gas. 

Most of the cold gas in the Universe is in the form of neutral hydrogen. While the molecular neutral hydrogen (\hj) is thought to serve as the fuel of star formation, the atomic neutral hydrogen (\hi) is indirectly connected to star formation through its conversion to \hj\ \citep[see e.g.,][]{Bigiel2008,Bigiel2010,Krumholz2012}. However, \hi\ in the local Universe is more abundant and much easier to observe with the 21~cm hyperfine emission line than \hj. It is therefore of great interest to explore the \hi\ reservoir through large 21~cm surveys, e.g., the \hi\ Parkes All-Sky Survey \citep[HIPASS;][]{Barnes2001,Meyer2004}, the Arecibo Fast Legacy ALFA Survey \citep[ALFALFA;][]{Giovanelli2005,Haynes2011,Haynes2018}, and the GALEX Arecibo SDSS Survey \citep[GASS;][]{Catinella2010,Catinella2018}. Along with the surveys measuring the \hj\ content through the CO emission lines \citep[e.g., the COLD GASS survey;][]{Saintonge2011,Saintonge2017}, we will be able to answer whether the galaxy quenching is due to the lack of \hi\ or the low conversion efficiency from \hi\ to \hj.

In order to explore the role of \hi\ in the galaxy quenching, it is necessary to measure the \hi\ masses for both the star-forming and quenched galaxies. However, individual source detection in blind \hi\ surveys like ALFALFA depends on the instrument flux limit \citep{Haynes2011}. Since quenched galaxies typically have low \hi\ fluxes, their detection rate in ALFALFA is therefore much smaller than that of the star-forming galaxies \citep{Huang2012}. Although the \hi\ detection rate still encodes useful information about the \hi\ reservoir in different populations \citep[see e.g.,][]{Catinella2013}, accurate average \hi\ masses can be directly measured by stacking all the \hi\ signals for quenched and star-forming galaxies separated in bins of different properties. 

Such an \hi\ spectral stacking technique has been applied to measure the \hi\ scaling relations with galaxy stellar mass, color, SFR and stellar surface density \citep[see e.g.,][]{Verheijen2007,Fabello2011,Gereb2015,Brown2015}. \cite{Saintonge2016} stacked the GASS and COLD GASS galaxy samples over the SFR-$M_\ast$ plane. They found that both the \hi\ and \hj\ fractions decrease for star-forming galaxies when their stellar masses increase, while the star formation efficiency and \hi-to-\hj\ ratios remain roughly constant, indicating that the decrease of the cold gas reservoir is the main cause of the galaxy quenching. \cite{Brown2017} stacked the overlap regions between Sloan Digital Sky Survey (SDSS) DR7 Main Galaxy Sample \citep{Albareti2017} and the spring area of the ALFALFA 70\% sample. They showed that the \hi\ depletion in satellite galaxies depends more on the halo mass than local density. 

Taking advantage of the completed ALFALFA 100\% sample, \cite{Guo2020} directly measured the total \hi\ mass in halos ranging from $10^{11}\msunh$ to $10^{14}\msunh$, by stacking the \hi\ signals of entire groups from SDSS DR7. They found that the total halo \hi\ gas depends on both the halo mass and halo richness, which confirms the finding of \cite{Guo2017} that the galaxy \hi\ gas is closely related to the halo accretion history. They proposed that the measured \hi\ reservoir is consistent with a three-phase scenario, i.e. smooth \hi\ accretion in halos of $\log(M_{\rm h}/\msunh)<11.8$, reduction of \hi\ supply in halos of $11.8<\log(M_{\rm h}/\msunh)<13$, and slow \hi\ growth by (dry) mergers in more massive halos. 

In observation, phenomenological classifications of galaxy quenching mechanisms have been proposed to separate the internal and external causes, e.g. the mass quenching and environment quenching (or halo quenching) \citep{Peng2010}. In this paper, we will explore the cause of galaxy quenching by extending the work of \cite{Guo2020} and stacking the \hi\ signals for star-forming and quenched galaxies in different stellar and halo mass bins, which naturally separates the contributions from mass and halo quenching. We also present an improved \hi\ mass map in the SFR-$M_\ast$ plane, based on the work of \cite{Saintonge2016}.  We will only focus on the \hi\ gas of the central galaxies, complementary to the study of \cite{Brown2017}.

The structure of the paper is as follows. We describe the galaxy samples and stacking method in \S\ref{sec:data}. We present the results in \S\ref{sec:stack1} and \S\ref{sec:stack2}. We summarize and discuss the results in \S\ref{sec:discussion} and \S\ref{sec:summary}. Throughout the paper, the halo mass is in units of $\msunh$, while the stellar and \hi\ masses are in units of $\msun$. We assume a Hubble constant of $h=0.7$.

\section{Data and Method}\label{sec:data}

\subsection{HI and Optical Samples}
We use the same galaxy samples as in \cite{Guo2020}, with the \hi\ data from the spring sky of ALFALFA 100\% complete catalog \citep{Haynes2018,Jones2018}, which encompasses most of the optical galaxy sample from SDSS DR7. We adopt the galaxy group catalog from \cite{Lim2017} for the identification of central and satellite galaxies, as well as the halo mass estimates, with the redshift range limited to $0.0025<z<0.06$. We refer the readers to \cite{Guo2020} for more details. The measurements of \hi\ mass functions of this catalog have been presented in \cite{Jones2020}. As a result, our final sample consists of $22,810$ central galaxies in the \hi\ spectral stacking. We note that this sample includes many isolated central galaxies in low-mass halos.

The $M_\ast$ and SFR measurements for the central galaxies are obtained from the GSWLC-2 catalog of \cite{Salim2018}, which used state-of-the-art UV/optical spectral energy distribution fitting techniques. We note that for galaxies below the star formation main sequence (SFMS), with a low specific SFR (sSFR, defined as ${\rm SFR}/M_\ast$), there are non-negligible measurement errors and their sSFR estimates can only be considered as upper limits \citep{Salim2016}, as these galaxies are truly quiescent. It does not affect our \hi\ stacking for the overall populations of star-forming and quenched galaxies. However, when we stack the \hi\ gas in the SFR-$M_\ast$ plane, we only focus on galaxies with $\log(\rm{sSFR/yr^{-1}})>-12$. As will be shown in the following sections, our conclusions are not affected if we use a more conservative cutoff of $\log(\rm{sSFR/yr^{-1}})>-11.5$ or even slightly higher sSFRs.

\begin{figure}
	\centering
	\includegraphics[width=0.45\textwidth]{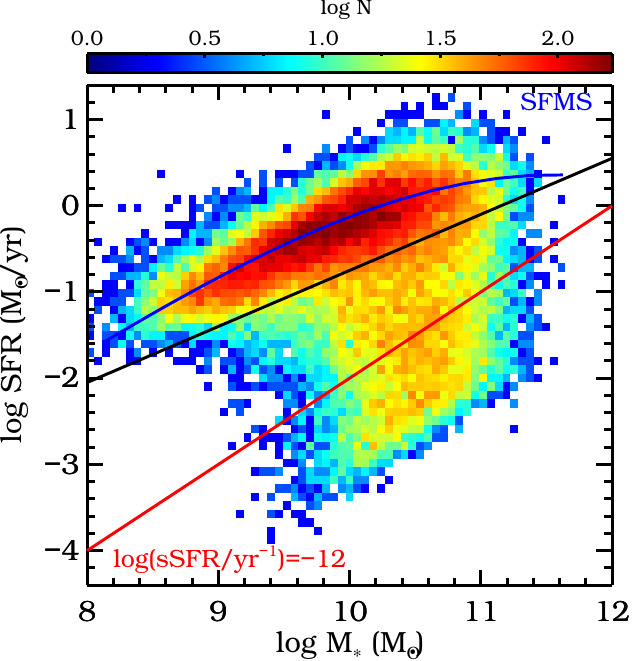}
	\caption{Distribution of the central galaxies as a function of SFR and $M_\ast$. The color scales represent the
		logarithmic number counts at the SFR and stellar mass intervals. The black line separates the star-forming and quenched populations, while the blue line is our definition of the SFMS and red line shows the positions of $\log(\rm{sSFR/yr^{-1}})=-12$. The color scale shows the
		logarithmic number counts.}
	\label{fig:sfr}
\end{figure}
We show in Figure~\ref{fig:sfr} the distribution of central galaxies from our sample in the SFR-$M_\ast$ plane. We fit a simple power law relation to classify the star-forming and quenched central galaxies as follows (shown as the black line),
\begin{equation}
	\log{\rm SFR_{cut}}=0.65\log M_\ast-7.25, \label{eq:sfcut}
\end{equation}
with the same power-law slope of 0.65 as that of \cite{Salim2007}. As the SFMS tends to be flattened for the massive galaxies \citep[e.g.,][]{Karim2011,Whitaker2012,Saintonge2016}, we follow the method of \cite{Saintonge2016} by fitting a third-order polynomial (shown as the blue line) to the mean SFRs of the star-forming galaxies as 
\begin{equation}
	\log{\rm SFR_{MS}}=-2.6121x+0.4605 x^2-0.0201 x^3, \label{eq:sfms}
\end{equation}
where $x=\log M_\ast$. For galaxies in the range of $9<\log(M_\ast/\msun)<10.5$, the SFMS can be well approximated by a power law of $\log{\rm SFR_{MS}}=0.65\log M_\ast-6.65$, which is also very similar to that of \cite{Janowiecki2020} using galaxies from the low-mass extensions of GASS \citep[xGASS;][]{Catinella2018}. 

As shown in Figure~\ref{fig:sfr}, this cut ensures that it is wide enough to include the majority of the main sequence galaxies in the star-forming population, as the typical star-forming galaxies are within $\pm0.3$~dex of the SFMS. The limit of reliable SFR measurements with $\log(\rm{sSFR/yr^{-1}})=-12$ is shown as the red line \citep{Salim2016}. There are still about 15\% of the galaxies below this limit.  

In addition, we also calculate the central stellar surface density within $1$~kpc, $\Sigma_1$, for these central galaxies to compare with other studies following the method introduced in \cite{Fang2013}. The central compactness $\Sigma_1$ is thought to be an important indicator of the galaxy quenching \citep{Cheung2012,Fang2013,Zolotov2015,Barro2017,WangEnci2018,Wang2020}. We use the surface brightness profiles in $ugriz$ bands from SDSS DR7 and have corrected for Galactic extinction. The $k$-corrections are applied in each annulus using version 4.2 of the \texttt{kcorrect} code from \cite{Blanton2007}. The corrected surface brightness profiles are then accumulated and interpolated smoothly to compute the total light within 1~kpc. Finally, the stellar mass surface density within 1~kpc is calculated using the total light within 1~kpc and $i$-band mass-to-light ratios calibrated by \cite{Fang2013}. 

\subsection{Stacking Method}\label{sec:method}
We follow the same stacking procedure as detailed in \cite{Guo2020}, based on the method outlined in \cite{Fabello2011}. In \cite{Guo2020}, the angular size of the square stacking aperture used to extract the spectrum of each central galaxy from the ALFALFA data cubes, was set as $\max(200\,\rm{kpc}/D_{\rm A}, 8\arcmin)$, where $D_{\rm A}$ is the the angular diameter distance of the galaxy. As that work was focused on the global properties of the groups the aperture selection was not optimal and led to significant contributions to the \hi \ flux from confusion, particularly for centrals in massive halos or with large distances. In this paper, we improve the method in \cite{Guo2020} by using a smaller and adaptive aperture size for each central galaxy, as detailed below. 

In order to create an adaptive aperture size a scaling between stellar mass and \hi \ diameter is needed as the majority of the stacking targets are undetected in ALFALFA. To do this we combine the $M_\ast$--$M_\mathrm{HI}$ relation of \citet{Huang2012} with the \hi \ size--mass relation \citep{Wang2016}. Here we note that for a stellar mass-selected sample of galaxies the average \hi \ mass (and therefore \hi \ diameter) is weakly dependent on the stellar mass \citep[e.g.][]{Catinella2010,Brown2015}. We therefore have intentionally constructed our relation based on the \hi-rich ALFALFA population in order to ensure apertures are large enough. We approximate the $M_\ast$--$M_\mathrm{HI}$ relation from \citet{Huang2012} for galaxies with $M_\ast > 10^{9.5} \mathrm{M_\odot}$ as a straight line: $\log(M_\mathrm{HI}/\mathrm{M_\odot}) = 0.256 \log(M_\ast/\mathrm{M_\odot}) + 7.263$. We then add a 0.5 dex offset to this relation so that it lies above most of the scatter in the ALFALFA population, again to ensure that apertures are large enough for even galaxies with exceptionally rich gas content. Combining this with the \citet{Wang2016} size--mass relation gives the final relation which we use to set our aperture widths: $\log(D_\mathrm{aperture}/\mathrm{kpc}) = 0.130 \log(M_\ast/\mathrm{M_\odot}) + 0.635$. 

We note that due to the resolution of Arecibo ($\sim$3.8\arcmin \ in L-band with ALFA) we limit the minimum aperture size to 4\arcmin. In practice the aperture sizes for the vast majority (almost 96\%) of the centrals in our catalogue are determined by this limit, and the relation above is only relevant for the centrals closer than $\sim$100 Mpc. It should also be noted that all apertures were extracted by rounding up to the next integer number of arcmin as an ALFALFA pixel is 1\arcmin$\times$1\arcmin \ and the extraction procedure does not allow partial pixels.

After all the extracted spectra in a particular stacking bin have been aligned in the rest frame and co-added, the spectrum is re-baselined following the same procedure as in \cite{Guo2020}. The \hi\ mass of the centrals in the stack is then finally estimated by integrating all emission within $\pm 300$ \kms. We confirm from the stacked spectra that there is not much emission outside this velocity range for our results in this paper, as shown similarly in Figure~B2 of \cite{Guo2020}. 

The statistical uncertainties in stacked \hi \ masses were estimated as $\sigma_\mathrm{rms}/\sqrt{N_\mathrm{chan}}$, that is, the rms noise in the stacked spectrum divided by the square root of the number of channels summed over. 
We found that the bootstrap uncertainty estimates from \cite{Guo2020} were well correlated with the statistical uncertainty estimates, with the bootstrap errors larger by a factor of around 1.5. For simplicity, we therefore multiply the statistical uncertainty estimates calculated here by 1.5 in order to mimic the bootstrap uncertainties estimated previously.

Following the prescriptions for estimating the contribution of confused \hi\ emission to a stack from \citet{Jones2016}, we estimate that at the highest redshift ($z=0.06$), where confusion is most severe, the expected \hi\ mass from confused emission in a stack of star-forming galaxies would be $\log(M_\mathrm{HI}/\mathrm{M_\odot})\sim8.8$, which is negligible compared to the expected \hi\ masses for those massive galaxies observed at these large distances. As confirmed by a careful investigation of stacking systematics using mock galaxy catalogs in \cite{Chauhan2021}, the choices of aperture size and velocity width in a reasonable range do not significantly affect the stacked \hi\ mass measurements.

\section{HI Stacking in the $M_\ast$-$M_{\rm h}$ Plane}\label{sec:stack1}
\subsection{Halo Mass Dependence}
\begin{figure}
	\centering
	\includegraphics[width=0.43\textwidth]{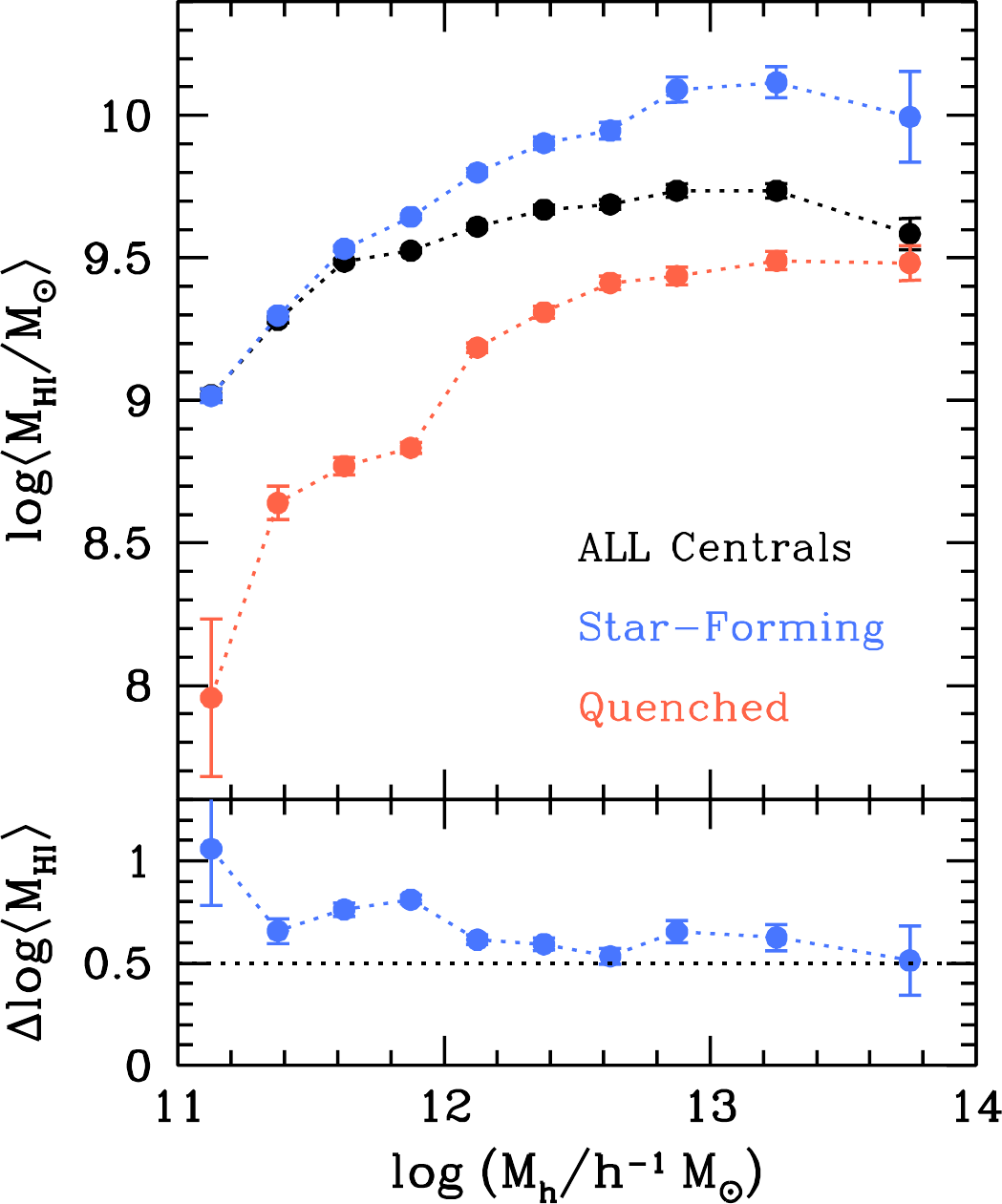}
	\caption{Top: measurements of \hi\ masses for central galaxies in halos of different masses. We show the measurements for all central galaxies (black), star-forming galaxies (blue) and quenched galaxies (orange). Bottom: offsets between the \hi\ masses from the star-forming and quenched galaxies, i.e. $\Delta\log M_{\rm HI}=\log M_{\rm HI, SFG}-\log M_{\rm HI, QG}$.}
	\label{fig:hihm}
\end{figure}
We show in the top panel of Figure~\ref{fig:hihm} the average \hi\ masses $\langle M_{\rm HI}\rangle$ of central galaxies in halos of different masses, for both the star-forming (blue line) and quenched (orange line) populations. The measurement for all centrals is also shown as the black line. The most straightforward conclusion is that in all halo mass environments, quenched galaxies (hereafter QGs) have much lower \hi\ gas content than the star-forming galaxies (hereafter SFGs). The \hi-halo mass relation for all the centrals can be simply understood as the combination of SFGs and QGs, along with the quenched fractions at different halo masses. 

For halos of $M_{\rm h}>10^{12}\msunh$, the \hi\ masses in QGs have a roughly constant lower offset of 0.6~dex (i.e., about four times smaller) compared to those of the SFGs, shown as the blue dashed line in the bottom panel of Figure~\ref{fig:hihm}. At lower halo masses, the offset becomes larger. It is remarkable to note that the shapes of the central \hi-halo mass relations for SFGs and QGs are quite similar. The \hi\ mass is smoothly increasing with the halo mass at the low mass end and becoming flat for halos of $M_{\rm h}>10^{12.5}\msunh$. For example, the \hi\ mass of SFGs increases by 1~dex from halos of $10^{11}\msunh$ to that of $10^{12.5}\msunh$, but there is almost no growth from $10^{12.5}\msunh$ to $10^{14}\msunh$. It implies that the smooth \hi\ accretion is impeded in massive halos of $M_{\rm h}>10^{12.5}\msunh$, irrespective of the star formation status of the central galaxies. It is potentially related to the effect of the virial shock-heating in these massive halos. As suggested by hydrodynamical simulations \citep{Birnboim2003,Keres2005}, gas falling into halos above a critical mass of around $10^{12}\msun$ will be shock-heated to the virial temperature and the hot accretion mode will then dominate the growth of massive galaxies. This is consistent with our finding of the flat gradient of the \hi-halo mass relation for the massive halos. 

\subsection{Joint Stellar and Halo Mass Dependence}
\begin{figure*}
	\centering
	\includegraphics[width=0.9\textwidth]{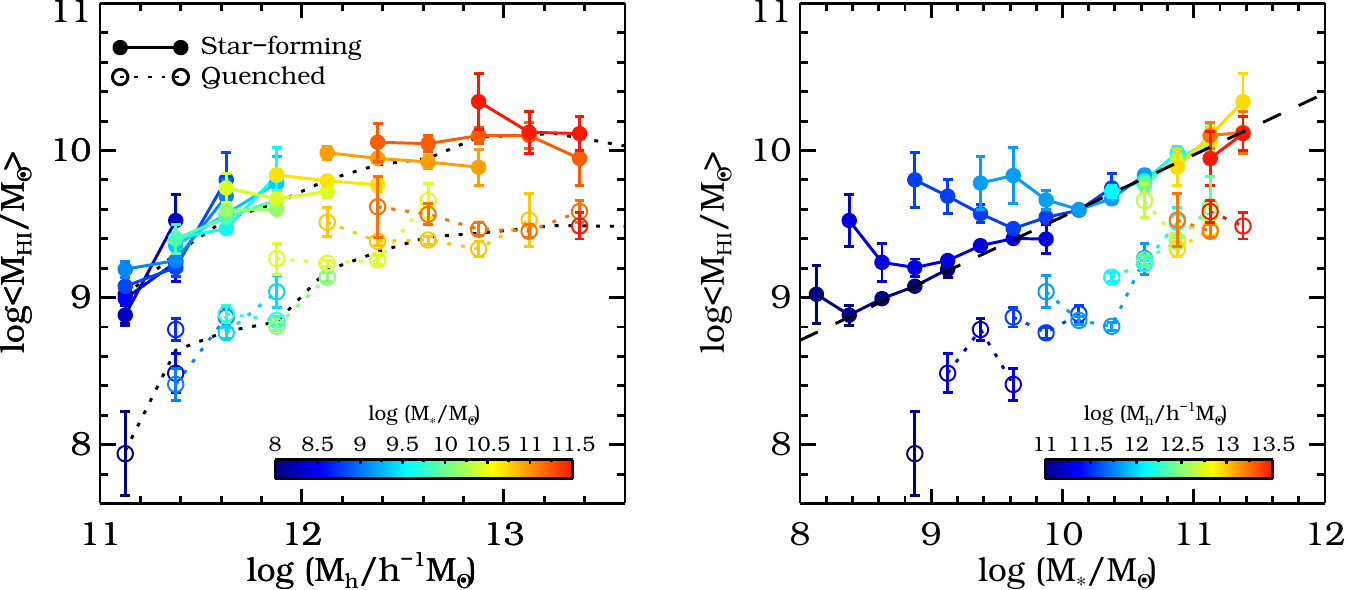}
	\caption{Measurements of mean \hi\ masses in different stellar and halo mass bins, i.e. $\langle M_{\rm HI}|M_\ast,M_{\rm h}\rangle$. The same set of data is shown in both the left and right panels, with the left showing the \hi-halo mass relation color-coded in $M_\ast$ and the right showing the \hi-stellar mass relation color-coded in $M_{\rm h}$. The SFGs are represented by the filled circles with solid lines, while the QGs are displayed with open circles and dotted lines. Measurements for galaxies in the same stellar (halo) mass bins are connected with lines in the left (right) panel. The two black dotted lines in the left panel show the \hi-halo mass relation for SFGs and QGs as in Figure~\ref{fig:hihm}. The black dashed line in the right panel represents our fit to the HIMS (see text). }
	\label{fig:hicen}
\end{figure*}
We show in Figure~\ref{fig:hicen} the \hi\ stacking in different stellar and halo mass bins, with $8<\log(M_\ast/\msun)<11.5$ and $11<\log(M_{\rm h}/\msunh)<13.5$, and the bin sizes are both set as 0.25~dex. The same measurements of $\langle M_{\rm HI}|M_\ast,M_{\rm h}\rangle$ are shown in both the left and right panels of Figure~\ref{fig:hicen}. The left panel shows the \hi-halo mass relation color-coded with $M_\ast$ and the right panel is the \hi-stellar mass relation color-coded with $M_{\rm h}$. The SFGs are represented by the filled circles with solid lines, while the QGs are displayed with open circles and dotted lines. Measurements for galaxies in the same stellar (halo) mass bins are connected with lines in the left (right) panel.

The measurements of \hi-halo mass relations in Figure~\ref{fig:hihm} are also shown as the black dotted lines for the SFGs and QGs in the left panel. By decomposing the contribution to the \hi\ signals from galaxies with different stellar masses, we find that for galaxies with $M_\ast>10^{10.5}\msun$, the dependence of $M_{\rm HI}$ on $M_{\rm h}$ at a given stellar mass bin is very weak, which is the same for both SFGs and QGs. The halo mass dependence seen in Figure~\ref{fig:hihm} for halos of $11.8<\log(M_{\rm h}/\msunh)<13.5$ just reflects the dependence on galaxy stellar mass. For galaxies with $M_\ast<10^{10.5}\msun$, the halo mass dependence is much stronger than that of the stellar mass.

It is clear that one effect of halo (or environment) quenching is to impede the accretion of cold gas, thus ensuring a long-term quenched state \citep{Dekel2014}. It takes effect for galaxies with a lower mass limit of $10^{10.5}\msun$, which roughly corresponds to a halo mass around $M_{\rm h}\sim10^{11.8}\msunh$ \citep[e.g.,][]{Moster2010,Yang2012}. However,  the roughly constant offset between the \hi\ masses of SFGs and QGs in all halo masses, indicates that the halo mass is not the direct cause of quenching.

From the \hi-stellar mass relation in the right panel of Figure~\ref{fig:hicen}, we find that SFGs tend to form a tight \hi\ main sequence (HIMS), which is fitted with a straight line as follows (shown as the black dashed line),
\begin{equation}
	\log M_{\rm HI,MS}=0.42\log M_\ast+5.35. \label{eq:hims}
\end{equation}
It is similar to the definition in \cite{Janowiecki2020}, with a slightly shallower slope. We note that for low mass halos, $M_{\rm HI}$ tends to slightly decrease with stellar mass before arriving at the HIMS. The resulting \hi\ fraction, $f_{\rm HI}\equiv M_{\rm HI}/M_\ast$, on the HIMS would scale as $f_{\rm HI}\propto M_\ast^{-0.58}$, i.e. the \hi\ fraction would decrease with increasing $M_\ast$ even for SFGs. It is consistent with the result of \cite{Saintonge2016}, with the flattening of the SFMS due to the overall reduction of the gas reservoirs.

For QGs, the behavior is still very similar. At a given halo mass, there is a very weak, if not negligible, dependence of $M_{\rm HI}$ on $M_\ast$ for $M_\ast<10^{10.5}\msun$. For more massive QGs, the $M_{\rm HI}$--$M_\ast$ relation has a similar slope as the HIMS, i.e. the \hi\ gas of QGs have the same relative growth rate $\dot{M}_{\rm HI}/M_{\rm HI}$ as that of SFGs. This remarkable similarity of the \hi-stellar mass relations for SFGs and QGs, again confirms that the stellar mass is also not the direct cause of quenching. 

Due to the tight correlation between central stellar and halo masses at the low-mass end \citep[e.g.,][]{Moster2010,Yang2012}, the average trend of halo mass dependence at the low-mass end can be reasonably reproduced when we combine the stellar mass dependence of $M_{\rm HI}\propto M_\ast^{0.42}$ with an average stellar-halo mass relation. This is not affected by the algorithm of the halo mass assignment in the group finder. For example, we can adopt the independent stellar-halo mass relation from \cite{Moster2010} as, 
\begin{equation}
	\langle M_\ast\rangle=0.056 M_{\rm h}\left[\left(\frac{M_{\rm h}}{M_1}\right)^{-1.07}+\left(\frac{M_{\rm h}}{M_1}\right)^{0.61}\right]^{-1},
\end{equation}
where $M_1=10^{11.9}\msun$. Therefore, given that $M_{\rm HI}\propto M_\ast^{0.42}$, the \hi-halo mass relation would scale as $M_{\rm HI}\propto M_{\rm h}^{0.87}$ and $M_{\rm HI}\propto M_{\rm h}^{0.16}$ at the low and high halo mass ends, respectively, consistent with our measurements. Despite the strong degeneracy between $M_\ast$ and $M_{\rm h}$, as shown in the left panel of Figure~\ref{fig:hicen}, for galaxies with $M_\ast$ below and above $10^{10.5}\msun$, $M_{\rm HI}$ is clearly dominated by the halo and stellar mass dependence, respectively. 

\subsection{Global Star Formation Law}\label{subsec:gsfl}
\begin{figure*}
	\centering
	\includegraphics[width=0.9\textwidth]{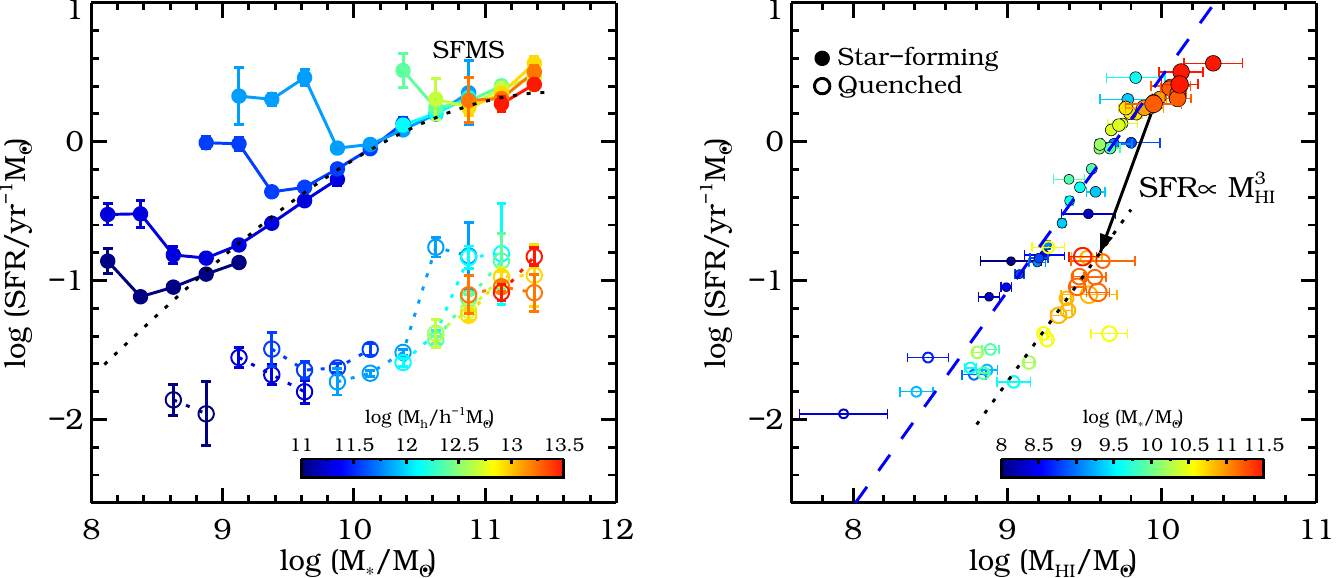}
	\caption{Left: similar to the right panel of Figure~\ref{fig:hicen}, but in the SFR-$M_\ast$ plane. The black dotted line shows the positions of the SFMS as in Figure~\ref{fig:sfr}. Right: the relation between SFR and $M_{\rm HI}$ for SFGs (filled circles) and QGs (open circles). The galaxy stellar mass is color-coded as displayed, and the symbol size represents the halo mass, with larger symbols for more massive halos. The power law relations of SFR $\propto M_{\rm HI}^{1.547}$ are shown as the blue dashed line and black dotted line for the SFGs and QGs, respectively. The black arrow indicates the example of ${\rm SFR}\propto M_{\rm HI}^3$ for galaxies with $M_\ast\sim10^{11}\msun$. For clarity, we only show the errors on $M_{\rm HI}$. }
	\label{fig:hisfr}
\end{figure*}
As the \hi\ gas provides the reservoir for star formation, we investigate the corresponding dependence of SFR on $M_\ast$ and $M_{\rm h}$ in the left panel of Figure~\ref{fig:hisfr}, which is similar to the right panel of Figure~\ref{fig:hicen}. We measure the mean SFR in the same $M_\ast$ and $M_{\rm h}$ bins as the \hi\ mass. The SFMS of Eq. (\ref{eq:sfms}) is denoted as the dotted black line. We find that the trends of SFR with $M_\ast$ and $M_{\rm h}$ are very similar to those of $M_{\rm HI}$ for both SFGs and QGs. As expected, most of the data points for SFGs lie on the SFMS, while there is similar halo mass dependence as seen in the \hi\ mass at the low mass end. The SFMS indeed tends to be flattened for $\log(M_\ast/\msun)>10.6$, similar to that of \cite{Saintonge2016}. We note that the gap between SFRs of SFGs and QGs is larger than that of the \hi\ mass.

We show, in the right panel of Figure~\ref{fig:hisfr}, the relation between SFR and $M_{\rm HI}$ for galaxies with different $M_\ast$ (color-coded) and different $M_{\rm h}$. Galaxies in more massive halos are shown in larger symbols. The filled and open circles are for the SFGs and QGs, respectively. As naturally expected from the scaling relations of SFMS and HIMS, the SFRs of SFGs would scale with $M_{\rm HI}$ as SFR $\propto M_{\rm HI}^\alpha$, with $\alpha=0.65/0.42=1.547$, shown as the blue dashed line. It is manifested in the tight distribution of SFGs with different $M_\ast$ and $M_{\rm h}$ along this line. At the massive end, the slope is slightly shallower than 1.547 due to the flattening of SFMS. This slope is close to the prediction of Kennicutt-Schmidt Law with $\alpha=1.4$ \citep{Schmidt1959,Kennicutt1998}, which is, however, based on the relation between SFR surface density and total cold gas surface density. 

The QGs with $M_\ast<10^{9.5}\msun$ still lie close to the bottom tail of the power law relation, as these galaxies are still not far from the SFMS, likely related to the selection effect when we get close to the detection limit of the SFR measurements. As will be discussed in the following sections, these galaxies are likely not fully quenched and would possibly return back to the star-forming population in their later evolution, accompanying with the cold gas accretion during the halo growth. For more massive galaxies, their SFR-$M_{\rm HI}$ relation can still be reasonably fitted with the same power law index of $\alpha=1.547$, shown as the black dotted line in the right panel of Figure~\ref{fig:hisfr}. It indicates that even for QGs, the SFR is still closely related to the available \hi\ mass, but the star formation efficiency (${\rm SFE}={\rm SFR}/M_{\rm HI}$) is significantly reduced. 

It is generally conceived that the SFR is more correlated with the molecular hydrogen \hj\ than \hi, with $\Sigma_{\rm SFR}\sim\Sigma_{\rm HI}^3$ \citep[e.g.,][]{Bigiel2008,Leroy2008}, where the SFR and $M_{\rm HI}$ are expressed in surface densities. It makes $M_{\rm HI}$ a poor indicator of SFR, which seems to contradict with our results above. However, it is simply because the variation of SFR in the relation ${\rm SFR}\propto M_{\rm HI}^{1.547}$ is mainly caused by the change of stellar mass, rather than the star formation status. If we compare the SFR-$M_{\rm HI}$ relations for galaxies with the same stellar mass, e.g. $M_\ast\sim10^{11}\msun$, the change of SFR from SFGs to QGs would roughly scale as ${\rm SFR}\propto M_{\rm HI}^3$ (shown as the black arrow in the right panel of Figure~\ref{fig:hisfr}), in very good agreement with \cite{Bigiel2008}. It can also be understood from the global relation among the three physical parameters of SFR, $M_{\rm HI}$ and $M_\ast$, as follows.

According to our measurements, we assume the following power law scaling relation,
\begin{equation}
	\log{\rm SFR}=\alpha\log M_{\rm HI} +\beta\log M_\ast+ \gamma, \label{eq:sfrhi}
\end{equation}
where $\gamma$ is a constant coefficient. We can estimate $\alpha$ and $\beta$ from Figure~\ref{fig:hisfr}. As shown in the left panel, at the massive end galaxies with the same stellar mass have a roughly constant offset of $\Delta\log{\rm SFR}\sim1.6$ from SFGs to QGs, and the corresponding \hi\ mass changes by $\Delta\log M_{\rm HI}\sim0.6$ as in Figure~\ref{fig:hihm}, which means that $\alpha\sim2.7$. In the right panel, at a given $M_{\rm HI}$, the vertical offset between the blue dashed and black dotted lines is $\Delta\log{\rm SFR}=-0.65$ while the stellar mass offset is around $\Delta\log M_\ast\sim1.25$, so $\beta\sim-0.52$. Therefore, we have ${\rm SFR}\propto M_{\rm HI}^{2.7}/ M_\ast^{0.52}$, consistent with our rough estimate before. At a given $M_{\rm HI}$, more massive galaxies would actually have lower SFRs, due to the reduction in gas fraction. It indicates that the SFR is determined by a global star formation law which depends on both the \hi\ reservoir and the stellar mass. 

\subsection{Atomic Gas Depletion Time}
\begin{figure*}
	\centering
	\includegraphics[width=0.9\textwidth]{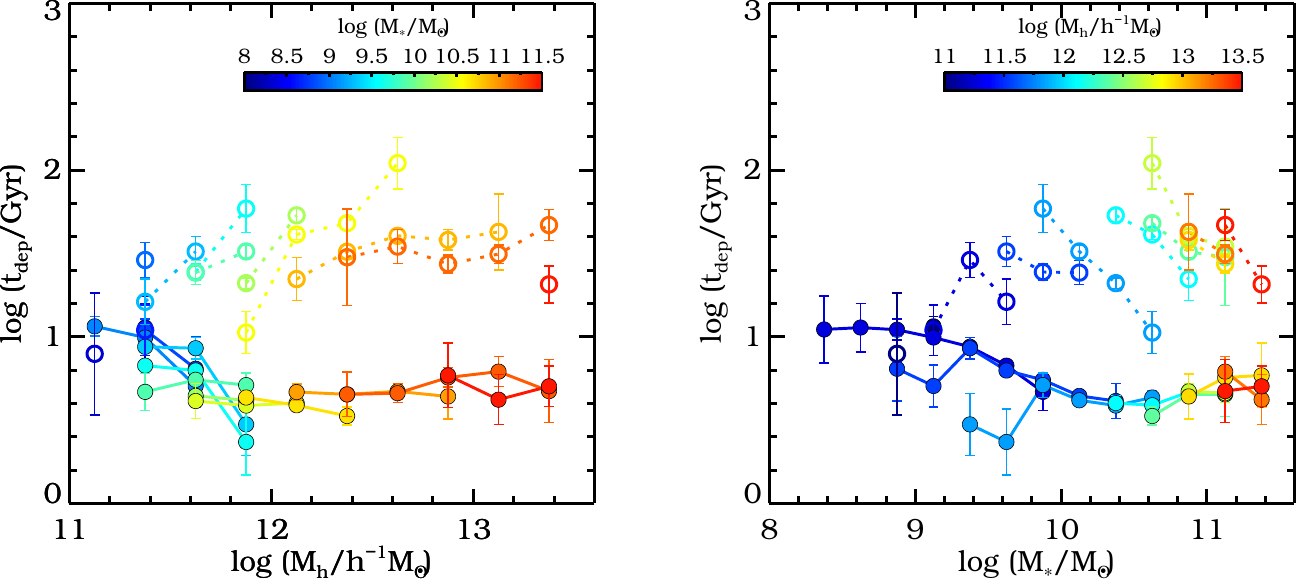}
	\caption{Similar to Figure~\ref{fig:hicen}, but for the atomic gas depletion time $t_{\rm dep}$. The errors for $t_{\rm dep}$ are calculated by adding in quadrature the errors of SFR and $M_{\rm HI}$.}
	\label{fig:tdep}
\end{figure*}
We define the atomic gas depletion time as, $t_{\rm dep}=M_{\rm HI}/{\rm SFR}$, which is the time scale for the galaxy to consume all the \hi\ gas with the current SFR. We show in Figure~\ref{fig:tdep} the dependence of $t_{\rm dep}$ on $M_{\rm h}$ and $M_\ast$ in the same format as in Figure~\ref{fig:hicen}. The SFGs have much smaller gas depletion time than those of the QGs, and the dependence on $M_{\rm h}$ and $M_\ast$ is very weak for massive galaxies with $M_\ast>10^{10.5}\msun$. \cite{Saintonge2017} found an average \hi\ gas depletion time of $\log(t_{\rm dep}/{\rm Gyr})$ around $0.65\pm0.44$ for galaxies in the SFMS using xGASS, which is consistent with our measurements. 

The weak dependence on $M_{\rm h}$ and $M_\ast$ is expected, as we find ${\rm SFR}\propto M_{\rm HI}^{1.5}$ for galaxies on the SFMS, which means that $t_{\rm dep}\propto M_{\rm HI}^{-0.5}\propto M_\ast^{-0.21}$. The increase of $t_{\rm dep}$ towards the low-mass end \citep[also seen in][]{Saintonge2017} is due to the different slope of the SFR-$M_\ast$ relation at a give halo mass bin compared to the SFMS (left panel of Figure~\ref{fig:hisfr}).

For QGs, $t_{\rm dep}$ is typically much larger than 10~Gyrs, indicating that the \hi\ gas is not actively involved in the star formation. However, the scaling relation as in Eq. (\ref{eq:sfrhi}) is still valid, as the \hi\ gas still provides the reservoir for star formation, though with very low efficiency. It may seem that the \hi\ reservoir is already large enough compared to the SFR of QGs and the quenching appears to be caused by the bottleneck of the conversion from \hi\ to \hj. However, as we show in Equation~(\ref{eq:sfrhi}), the star formation efficiency would be significantly decreased with the decreasing of \hi\ reservoir when $\alpha$ is larger than unity. 
There is positive dependence on $M_{\rm h}$ and negative dependence on $M_\ast$ for low mass QGs below $10^{10.5}\msun$, due to the similar behavior of \hi\ mass as in Figure~\ref{fig:hicen}. 

As will be further confirmed in the next section, both the SFGs and QGs follow very similar global star formation law of Equation~(\ref{eq:sfrhi}). It implies that galaxies at low redshifts tend to live in a quasi-steady state in which the star formation is directly regulated through the fuelling and depletion of the cold gas \citep{Bouche2010,Lilly2013,Dekel2014b}.  

\section{HI Stacking in the SFR-$M_\ast$ Plane}\label{sec:stack2}
We show in the above sections the \hi\ stacking in different stellar and halo mass bins, for the SFGs and QGs. To fully investigate the smooth evolution of \hi\ gas in the transition from SFGs to QGs, we show in this section the stacking  in bins of SFR and $M_\ast$, with $8<\log(M_\ast/\msun)<11.5$ and $-3.5<\log{\rm SFR}<1.5$ and the bin sizes are 0.25~dex and 0.5~dex, respectively.

\subsection{Dependence on SFR and $M_\ast$}
\begin{figure*}
	\centering
	\includegraphics[width=0.9\textwidth]{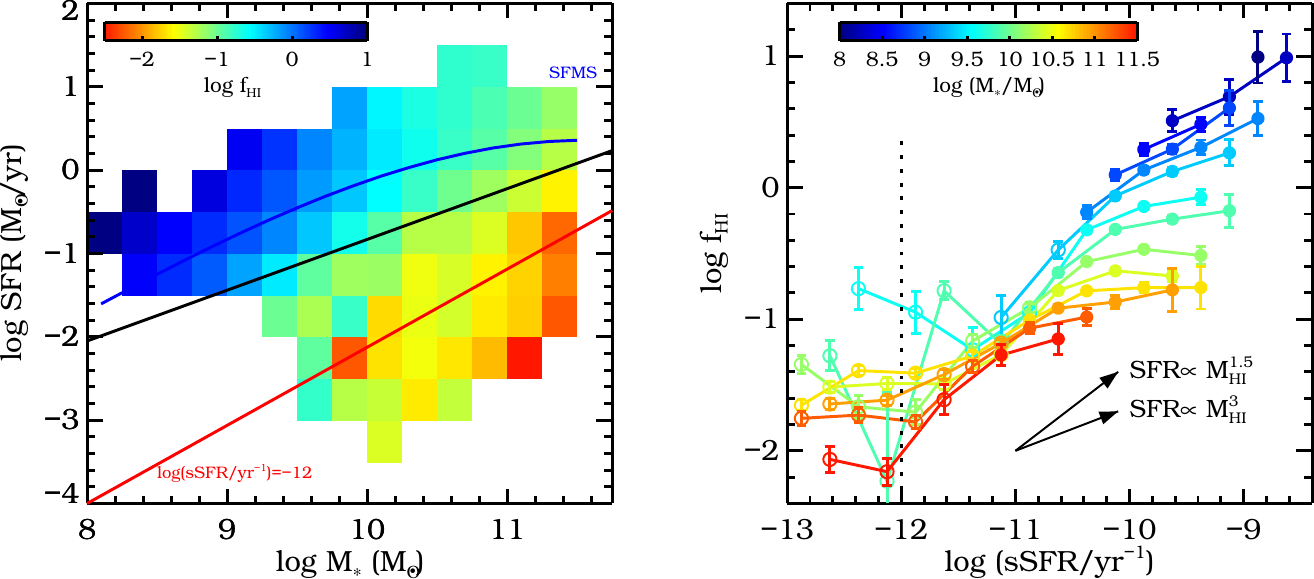}
	\caption{Left: color-coded map of $f_{\rm HI}$ in different SFR and $M_\ast$ bins. The lines of SFMS, separation between SFGs and QGs, as well as the limit of $\log({\rm sSFR/yr^{-1}})=-12$ are also shown, the same as in Figure~\ref{fig:sfr}. Right: measurements of $f_{\rm HI}$ as a function of sSFR in different stellar mass bins. The filled and open circles are for SFGs and QGs, respectively. The Measurements are only reliable for galaxies with $\log({\rm sSFR/yr^{-1}})>-12$.}
	\label{fig:hisfrsm}
\end{figure*}
We show in the left panel of Figure~\ref{fig:hisfrsm} the stacked \hi\ signals as a function of SFR and $M_\ast$. For fair comparisons with literature, we show here the \hi\ gas fraction defined as $f_{\rm HI}\equiv M_{\rm HI}/M_\ast$, while the original stacking is still made for $M_{\rm HI}$ rather than $f_{\rm HI}$. The lines of SFMS, demarcation line between SFGs and QGs, as well as the limit of $\log({\rm sSFR/yr^{-1}})=-12$ are also shown, the same as in Figure~\ref{fig:sfr}. We note that the stacking measurements for galaxies with $\log({\rm sSFR/yr^{-1}})<-12$ are less reliable due to the large uncertainties of SFR. But we still show them in the figure for consistency.

There are two apparent trends in this figure. For SFGs, $f_{\rm HI}$ is decreasing with increasing $M_{\ast}$ along the main sequence, consistent with the finding of \cite{Saintonge2016}. For QGs, most of their \hi\ fractions are below 0.1, and $f_{\rm HI}$ is continually smaller when the galaxies are farther away from the SFMS. It further confirms our previous conclusion that galaxy quenching is directly caused by the decreasing of the \hi\ gas content. This is also seen in the right panel of Figure~\ref{fig:hisfrsm}, where we show the dependence of $f_{\rm HI}$ on sSFR for different stellar masses. The filled and open circles are for SFGs and QGs, respectively. Lines of different colors are for galaxies with different $M_\ast$.

For SFGs, there is significant variation of $f_{\rm HI}$ with sSFR, depending on the stellar mass. More massive galaxies usually have smaller $f_{\rm HI}$, as seen in the left panel. However, the QGs with $\log({\rm sSFR/yr^{-1}})>-12$ have a much tighter relation between $f_{\rm HI}$ and sSFR, with only very weak dependence on $M_\ast$. The other important systematic trend is that the slope of $d\log f_{\rm HI}/d\log{\rm sSFR}$ for high sSFRs varies from $1/1.5$ for low-mass galaxies to $0$ for massive ones, as indicated in the figure. The slopes all converge to around $1/1.5$ for QGs of different stellar masses. If we compare the slope from the SFGs to QGs at a given $M_\ast$, it is consistent with our previous estimate of a scaling relation of ${\rm SFR}\propto M_{\rm HI}^3$ in Section~\ref{subsec:gsfl}. 

The relation between $f_{\rm HI}$ and sSFR is potentially related to the star formation law, although our measurements are for globally averaged galaxies and the optical star formation radii may be much smaller than the \hi\ disk sizes. With resolved measurements in the inner parts of nearby galaxies, \cite{Bigiel2008} found a very similar change of slope in the relation $\Sigma_{\rm SFR}\propto\Sigma_{\rm HI}^\alpha$, with $\alpha$ varying from around $1.5$ to infinity when $\Sigma_{\rm HI}$ increases. \cite{Bigiel2010} found a slope of $\alpha\sim1.7$ for the outer disks of the galaxies, where the star formation efficiency is much lower than the inner parts. It is similar to the situation of the QGs, with significantly decreased \hi\ surface densities compared to those of SFGs with the same masses. 

\begin{figure}
	\centering
	\includegraphics[width=0.45\textwidth]{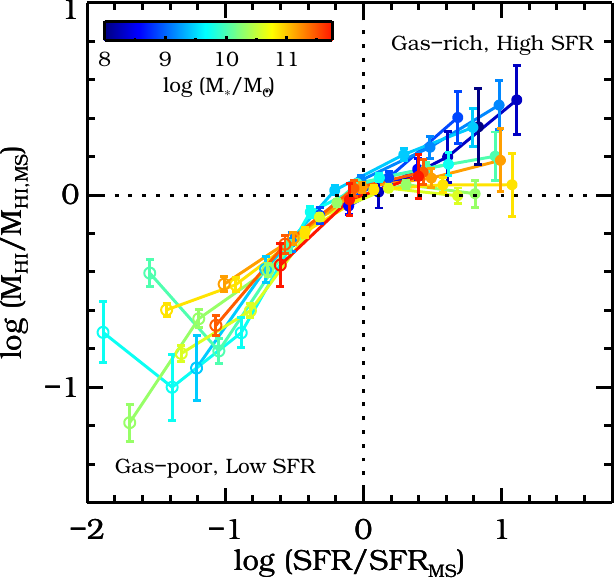}
	\caption{Similar to the right panel of Figure~\ref{fig:hisfrsm}, but we scale the SFR and $M_{\rm HI}$ by the corresponding values at the SFMS, i.e. ${\rm SFR/SFR_{MS}}$ and $M_{\rm HI}/M_{\rm HI,MS}$. The horizontal and vertical dotted lines represent the positions of $M_{\rm HI}=M_{\rm HI,MS}$ and ${\rm SFR=SFR_{MS}}$, respectively. Measurements are only shown for $\log({\rm sSFR/yr^{-1}})>-12$.}
	\label{fig:hims}
\end{figure}
It is more informative if we scale the SFR and $M_{\rm HI}$ by the corresponding values at the SFMS (Figures~\ref{fig:hicen} and~\ref{fig:hisfr}), i.e. $\Delta\log{\rm SFR}\equiv\log({\rm SFR/SFR_{MS}})$ and $\Delta\log M_{\rm HI}\equiv \log(M_{\rm HI}/M_{\rm HI,MS})$. As shown in Figure~\ref{fig:hims}, the galaxies drop below the SFMS when their \hi\ masses decrease from the HIMS. While the SFGs have the apparent trend of increasing slope of $\alpha$ with $M_\ast$, they drop below the SFMS at almost the same rate with $\alpha\sim1.5$, forming a very tight relation between SFR and $M_{\rm HI}$ irrespective of the stellar mass. For massive SFGs, the significantly increased slope $\alpha$ is likely caused by two reasons. Firstly, those most massive galaxies with high SFRs tend to efficiently consume the \hi\ reservoir and form the molecular gas \citep{Catinella2018}. The other is that the \hi\ accretion in these systems is slowed down, as manifested by the significantly decreased $f_{\rm HI}$. These two effects make the slope between SFR and $M_{\rm HI}$ much flatter, consistent with the results of \cite{Jaskot2015}. This overall trend further confirms our previous conclusion that the central galaxies are quenched by the overall reduction of the \hi\ gas reservoir by an amount of $\sim0.6$--$1.0$~dex. 

Taking advantage of the stacking method, we are able to quantitatively trace the change of the relation between SFR and $M_{\rm HI}$ above and below the SFMS, improving the similar investigation shown in \cite{Janowiecki2020} using the xGASS sample (their Figure~5). However, our measurements for $\Delta\log{\rm SFR}<-1$ suffer from large uncertainties of the SFR measurements, making it hard to predict the slope $\alpha$ for these truly quiescent galaxies. 

\subsection{Connection to the Central Stellar Surface Density}
\begin{figure}
	\centering
	\includegraphics[width=0.48\textwidth]{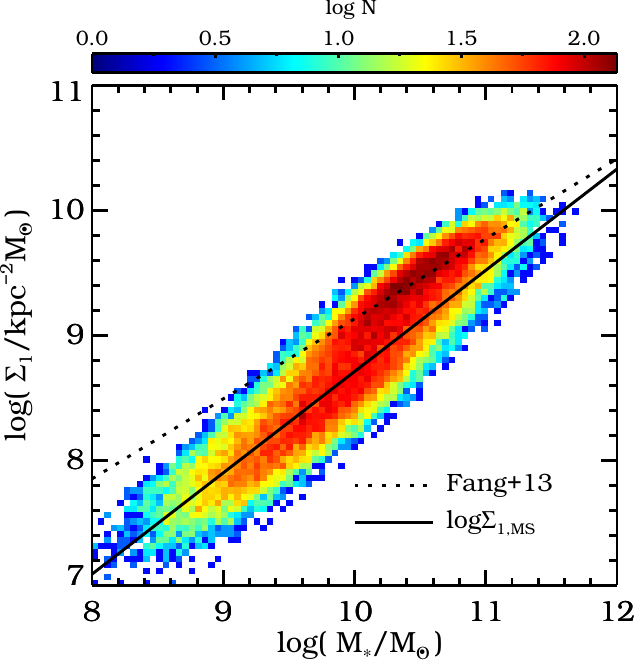}
	\caption{Distribution of $\Sigma_1$ as a function of $M_\ast$ in our sample, color-coded in the logarithmic number counts. We estimate the average of $\Sigma_1$ on the main sequence ($\Sigma_{\rm 1,MS}$) using only the SFGs, shown as the solid line. For comparison, the ridge line for quenched galaxies from \cite{Fang2013} ($\log\Sigma_1=0.64\log M_\ast+2.73$) is also shown as the dotted line.}
	\label{fig:sigmasm}
\end{figure}
\begin{figure*}
	\centering
	\includegraphics[width=0.9\textwidth]{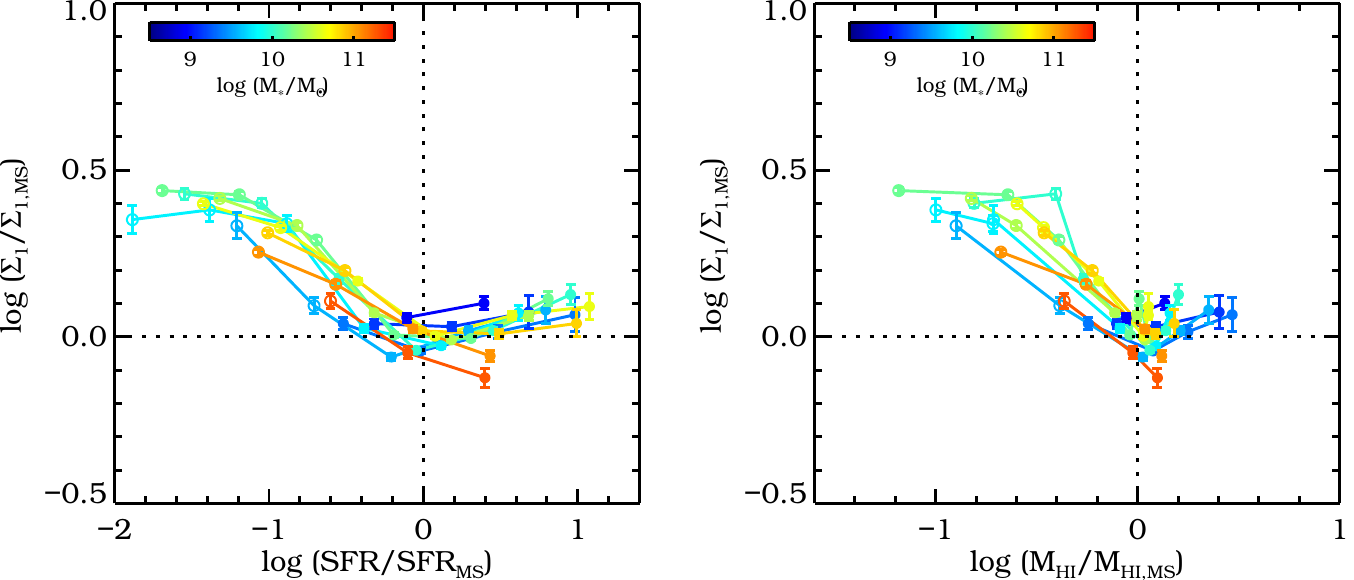}
	\caption{Dependence of $\Sigma_1$ on the SFR (left panel) and $M_{\rm HI}$ (right panel) in different stellar mass bins (color coded). The filled and open circles are for the SFGs and QGs, respectively. The positions of the zero points are labeled with the dotted lines. The measurements of $\Sigma_1$, SFR and $M_{\rm HI}$ are all scaled by their values on the main sequence.}
	\label{fig:hisigma1}
\end{figure*}
Increasing evidence of the structural differences between the SFGs and QGs suggests that the formation of a dense core (represented by the measurement of $\Sigma_1$) is the prerequisite for galaxy quenching \citep[see e.g.,][]{Kauffmann2003,Schiminovich2007,Fang2013,Barro2013,vanDokkum2014}. We show in Figure~\ref{fig:sigmasm} the distribution of $\Sigma_1$ as a function of $M_\ast$ in our sample. We also fit the average of $\Sigma_1$ on the main sequence ($\Sigma_{\rm 1,MS}$) using only the SFGs as,
\begin{equation}
	\log\Sigma_{\rm 1,MS}=0.810\log M_\ast+0.607 \label{eq:sigms}
\end{equation}
which is shown as the solid line in the figure. The ridge line of quenched galaxies from \cite{Fang2013} ($\log\Sigma_1=0.64\log M_\ast+2.73$) is displayed as the dotted line. More massive galaxies have significantly denser cores and the QGs have even higher $\Sigma_1$ values. We focus on the measurements of $\Sigma_1$ for $M_\ast>10^{8.5}\msun$ to ensure enough S/N. 

We show in Figure~\ref{fig:hisigma1} the dependence of $\Sigma_1$ on SFR and $M_{\rm HI}$ in different stellar mass bins. All the measurements are scaled by their values on the main sequence, with the additional definition of $\Delta\Sigma_1\equiv\log(\Sigma_1/\Sigma_{\rm 1,MS})$. There are strong correlations among these measurements. We find similar trend between $\Delta\Sigma_1$ and $\Delta\log{\rm SFR}$ as in literature. The SFGs evolve along with the $\Sigma_{\rm 1,MS}$ ($\Delta\Sigma_1\sim0$) when their SFRs gradually fall on the main sequence. As the galaxies leave the SFMS with $\Delta\log{\rm SFR}<0$, $\Delta\Sigma_1$ also increases until a plateau of $\Delta\Sigma_1\sim0.4$, which is the so-called compaction process \citep{Dekel2014}. The sharp decrease of SFR with $\Delta\Sigma_1$ remaining roughly constant identifies the final state of quenching. For our definition of QGs with $\Delta\log{\rm SFR}\leq-0.6$ as in Figure~\ref{fig:sfr}, the quenching happens along with the compaction.

The $\Delta\Sigma_1$--$\Delta\log M_{\rm HI}$ relation appears quite different for the SFGs compared to that of the QGs, as the majority of the SFGs locate around $\Delta\Sigma_1\sim0$ and $\Delta\log M_{\rm HI}\sim0$. It indicates that the \hi\ gas increases consistently with central core density. As we have seen in Figure~\ref{fig:hims}, most massive SFGs lie very close to the HIMS. The similar behavior is seen for $\Sigma_1$, i.e. as galaxies evolve along the $\Sigma_1$ main sequence, their HI masses also follow the HIMS closely. The trend of $\Delta\Sigma_1$ with $\Delta\log M_{\rm HI}$ for QGs show very similar slopes for different stellar masses, with $\Sigma_1\propto M_{\rm HI}^{-0.5}$. 

We note that the scale lengths of these three parameters are quite different, ranging from $1$~kpc to a few tens of kpc for the \hi\ disks. The tight correlations among their values implies that the same physical mechanisms are regulating their joint evolution. Our results are in good agreement with the compaction evolutionary scenario suggested in theory and simulations \citep[e.g.,][]{Dekel2014,Zolotov2015,Tacchella2016}. The dense SFGs (also called blue nuggets) could be formed from the wet (gas-rich) compaction induced by intense cold gas inflow due to dissipative processes such as mergers and violent disc instability, which is consistent with our finding in the right panel of Figure~\ref{fig:hisigma1}. The onset of galaxy quenching is typically associated with the maximum gas compactness in hydrodynamical simulations \citep[e.g.,][]{Zolotov2015}. We find that as the compaction starts, the \hi\ reservoir is also decreased along with $\Sigma_1$ potentially due to the gas consumption, outflows, and stellar or AGN feedback. In this model, the process of galaxies moving towards a compaction is relatively gradual and the drop of central gas density and SFR following the maximum of compaction is relatively quick. The increased $\Sigma_1$ away from the HIMS is consistent with such a scenario.

\begin{figure}
	\centering
	\includegraphics[width=0.43\textwidth]{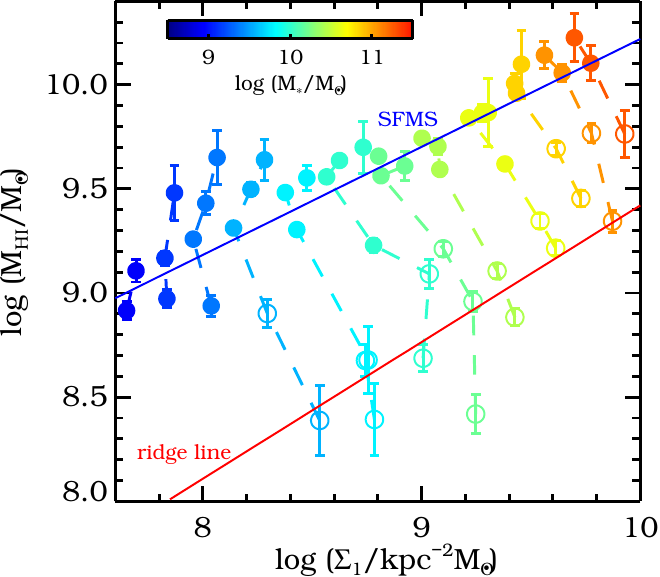}
	\caption{Relation between $M_{\rm HI}$ and $\Sigma_1$ for galaxies in different $M_\ast$ and SFR bins. The filled and open circles are for the SFGs and QGs, respectively. The galaxy stellar masses are color-coded. The solid blue line shows the positions of the HIMS and $\Sigma_1$ main sequence at different stellar masses as in Figures~\ref{fig:hicen} and~\ref{fig:sigmasm}. The solid red line indicates the corresponding positions of using $\Delta\log M_{\rm HI}=-0.7$ and the $\Sigma_1$ ridge line of \cite{Fang2013}.}
	\label{fig:hisigma2}
\end{figure}
We show in Figure~\ref{fig:hisigma2} the dependence of $M_{\rm HI}$ on $\Sigma_1$ for galaxies in different $M_\ast$ and SFR bins. Galaxies in the same $M_\ast$ bins are connected with dashed lines of different colors. The positions of the main sequences of $M_{\rm HI}$ and $\Sigma_1$ at different stellar masses are shown as the solid blue line. As expected, there is a tight relation between $M_{\rm HI}$ and $\Sigma_1$ for the SFGs, due to their strong dependence on $M_\ast$.  It is interesting that the quenching routes for galaxies with different $M_\ast$ are all perpendicular to the main sequence, with a slope of $M_{\rm HI}\propto \Sigma_1^{-2}$. The different $\Sigma_1$ values for QGs suggest that there is not a fixed quenching threshold in $\Sigma_1$ and it increases with stellar mass \citep{Barro2017}. 

While we do not have accurate measurements for galaxies with $\log({\rm sSFR/yr^{-1}})<-12$, it is not clear whether they would form another tight quiescent sequence. But as predicted from the hydrodynamical simulations \citep{Zolotov2015}, $\Sigma_1$ will remain roughly constant after the quenched cores are formed (the so-called red nugget phase). The \hi\ mass would possibly further decrease with the SFR until sSFR drops to an extremely low value, where a very minor amount of the \hi\ reservoir is involved in the star formation activity. To guide the eye, we show the red solid line as assuming $\Delta\log M_{\rm HI}=-0.7$ (corresponding to $\Delta\log{\rm SFR}\sim-1.2$) and the $\Sigma_1$ ridge line of \cite{Fang2013}. It roughly marks the positions where the quenched cores are formed. The trend of quenching is seen in some stellar mass bins in Figure~\ref{fig:hisigma2}, with the sharp drop of $M_{\rm HI}$ at roughly constant $\Sigma_1$ in the QGs.

\subsection{Dependence on Morphology}

The structural evolution from the SFGs to QGs usually accompanies the change of morphology from disk-dominated to bulge-dominated systems. The morphological quenching has been proposed in \cite{Martig2009}, emphasizing the effect of bulge in stabilizing the gas disk and preventing the cold gas from collapsing and forming stars. While such a mechanism could be contributing in some bulge-dominated galaxies, we find that the major reason for the central quenching is the depletion of \hi\ reservoir, which also increases the Toomre instability parameter Q \citep{Toomre1964} in stabilizing the disk by decreasing the cold gas surface density \citep{Dekel2014}.

\begin{figure*}
	\centering
	\includegraphics[width=0.9\textwidth]{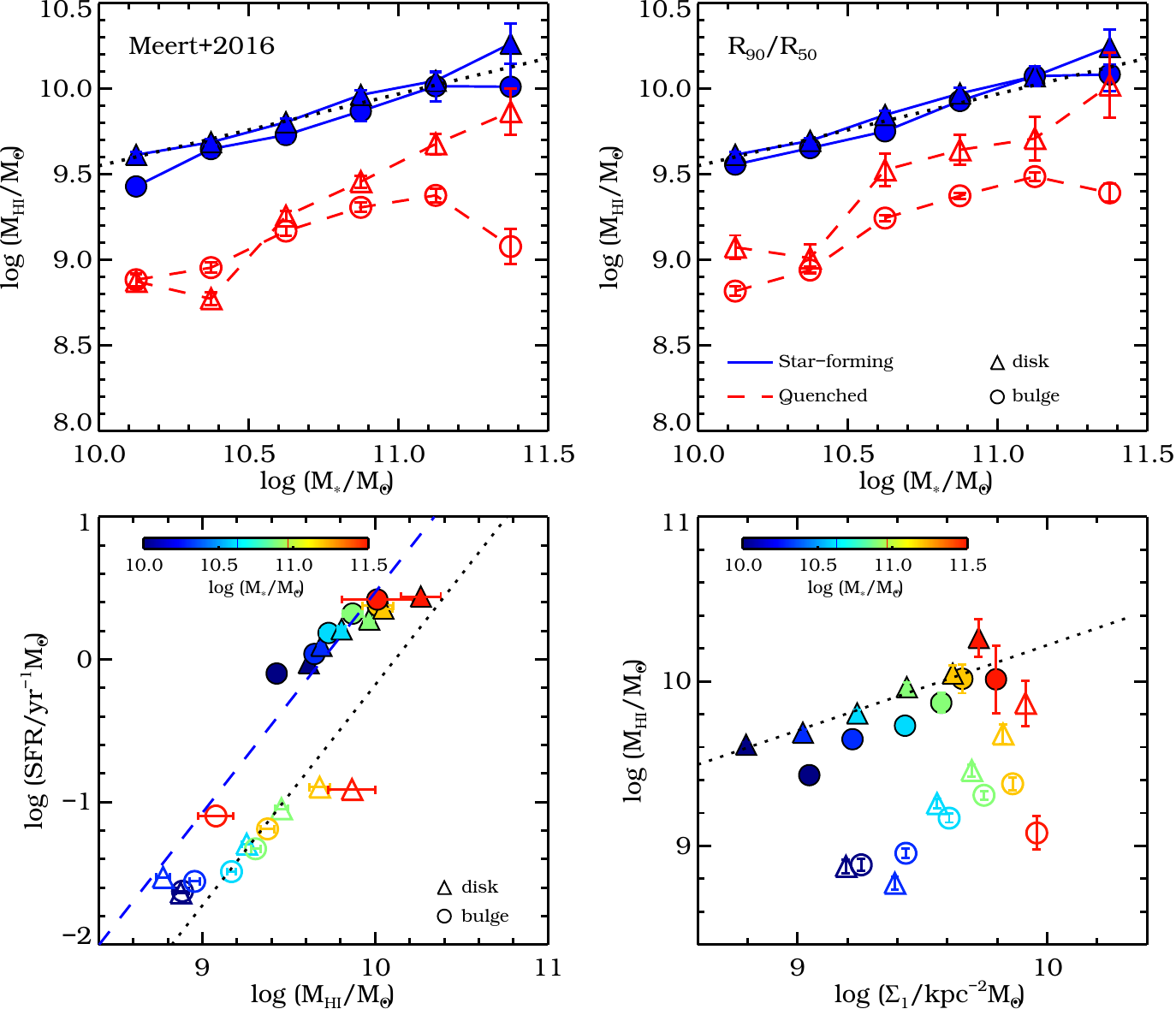}
	\caption{Top panels: measurements of $M_{\rm HI}$ for galaxies in four categories of star-forming/quenched disk and bulge-dominated systems. The results are shown for the morphology classifications of \cite{Meert2016} (top left) and $R_{90}/R_{50}$ (top right). The filled symbols with solid blue lines are for the SFGs, and the open symbols with red dashed lines are for the QGs. The disk- and bulge-dominated galaxies are displayed with triangles and circles, respectively. The black dotted lines in both panels represent the HIMS as in Figure~\ref{fig:hicen}. Bottom panels: distributions of star-forming/quenched disk- and bulge-dominated galaxies in the SFR--$M_{\rm HI}$ (bottom left) and $M_{\rm HI}$--$\Sigma_1$ diagrams (bottom right). The symbols are the same as in the top panels. We only show the results for the morphology classification of \cite{Meert2016}. The blue dashed and black dotted lines in the bottom left panel are the power law relations of ${\rm SFR}\propto M_{\rm HI}^{1.547}$ for the SFGs and QGs, respectively, as also shown in the right panel of Figure~\ref{fig:hisfr}. The dotted line in the bottom right panel shows the main sequence of $M_{\rm HI}$ and $\Sigma_1$, as in Figure~\ref{fig:hisigma2}. }
	\label{fig:hisfrmorph}
\end{figure*}
Galaxies can be generally separated into four categories, according to the morphology (disk or bulge dominated) and the star formation status (star-forming or quenched). While star-forming disk galaxies and quenched spheroidal galaxies are typically observed, the star-forming bulge-dominated and quenched disk galaxies are much less abundant. The quenched disk galaxies or red spiral galaxies have raised lots of interests recently \citep[see e.g.,][]{Masters2010,Cortese2012,Tojeiro2013,Schawinski2014,Bremer2018,Hao2019,GuoRui2020,LuoYu2020,Mahajan2020,Zhou2020}.

Quantifying the \hi\ reservoir for these quenched disk galaxies is a key step towards understanding their formation history. However, whether these galaxies have enough gas reservoir to support their star formation is still in debate. \cite{Zhang2019} found a surprisingly large amount of cold \hi\ reservoir for nearly all massive quenched disk central galaxies ($10.6<\log(M_\ast/\msun)<11$), comparable to that expected for their star-forming counterparts. \cite{Cortese2020} argued that the underestimation of the aperture-corrected SFR from the MPA/JHU SDSS DR7 catalog, as used in \cite{Zhang2019}, will wrongly classify a significant fraction of star-forming galaxies as quenched. 

To complement our understanding of the quenching processes along with the change of morphology, we also investigate the \hi\ reservoir for the star-forming/quenched disk and bulge-dominated central galaxies. We stack the \hi\ signals for the four categories of galaxies in our sample. We adopt two methods of morphology classifications. One is the 2D photometric two-component bulge/disk decomposition of \cite{Meert2016} for the SDSS DR7 galaxies. The catalog galaxies flagged as low quality are discarded. We use $0.5$ as the cut for the bulge-to-total light ratio (B/T) to divide the bulge- and disk-dominated galaxies. The other is the concentration index (defined as $R_{90}/R_{50}$ of the Petrosian flux) from the SDSS pipeline, which does not suffer from the uncertainties of the disk/bulge decomposition. As shown in \cite{Luo2020}, the cut of $R_{90}/R_{50}=2.5$ provides a reasonable demarcation line for the classic and pseudo bulges, as well as no-bulge and elliptical galaxies. We adopt the same cut of $R_{90}/R_{50}$ for our sample. A similar cut of $R_{90}/R_{50}=2.6$ was used in the \hi\ stacking of \cite{Fabello2011}.

To ensure enough S/N, we only stack the \hi\ signals for the stellar mass range of $10<\log(M_\ast/\msun)<11.5$ for the four populations. We show in the top panels of Figure~\ref{fig:hisfrmorph} the $M_{\rm HI}$ measurements for the morphology classifications of \cite{Meert2016} (left panel) and $R_{90}/R_{50}$ (right panel). The filled symbols with solid blue lines are for the SFGs, and the open symbols with red dashed lines are for the QGs. The disk- and bulge-dominated galaxies are displayed with triangles and circles, respectively. The black dotted lines represent the HIMS as in Figure~\ref{fig:hicen}. 

We note that the results from the two different morphology classifications are in very good agreement with each other, with only minor differences in the quenched disk galaxies. For SFGs, there is only weak dependence of $M_{\rm HI}$ on morphology, with bulge-dominated galaxies having slightly smaller \hi\ reservoir. The fact that the bulge- and disk-dominated galaxies lie close to each other is consistent with the tight HIMS seen in the whole sample. It also implies that the morphological transition during the star-forming phase, e.g. through compaction, does not significantly affect the overall \hi\ reservoir. This is in very good agreement with the results from \cite{Cook2019} using an independent bulge/disk decomposition method for the xGASS sample \citep[see also][]{Cook2020}.

For QGs, the differences in $M_{\rm HI}$ between disk- and bulge-dominated galaxies increases with $M_\ast$, with the more massive quenched disk galaxies approaching their star-forming counterparts. As will be shown below, the \hi\ masses conform to their SFRs, with quenched disk galaxies having slightly higher SFRs. However, for the mass range studied in \cite{Zhang2019} ($10.6<\log(M_\ast/\msun)<11$), the quenched disk galaxies have on average 0.35--0.5~dex lower \hi\ masses compared to the star-forming disks, depending on the morphology classification methods. Even for more massive quenched disk galaxies, they have consistently lower \hi\ fractions relative to the star-forming ones. This again suggests that central galaxies are quenched by the depletion of the \hi\ reservoir, even for the disk galaxies.

But the large difference seen in the quenched disk and bulge galaxies is still intriguing. It is helpful to compare their distributions in the SFR--$M_{\rm HI}$ and $M_{\rm HI}$--$\Sigma_1$ diagrams, as shown in the bottom panels of Figure~\ref{fig:hisfrmorph}, where the symbols are the same as in the top panels. We only show the results for the morphology classification of \cite{Meert2016}. The results of using $R_{90}/R_{50}$ are quite similar. The blue dashed and black dotted lines in the left panel are the power law relations of ${\rm SFR}\propto M_{\rm HI}^{1.547}$ for the SFGs and QGs, respectively, as also shown in the right panel of Figure~\ref{fig:hisfr}.

We find that there is no significant difference between the global star formation scaling relations between the disk- and bulge-dominated galaxies in the star-forming and quenched populations. It means that their \hi\ masses are consistent with their observed SFRs and the global star formation law of Equation~(\ref{eq:sfrhi}) applies to both the disk and bulge-dominated galaxies.

We note that the quenched disk galaxies all have higher $\Sigma_1$ than star-forming disk and bulge galaxies with the same $M_\ast$. It indicates that these quenched disk galaxies have already undergone the compaction processes, with a strong central bulge. There is no substantial difference in the \hi\ mass for the quenched disk and bulge galaxies with $M_\ast<10^{11}\msun$. As also found in a detailed study of \cite{GuoRui2020}, the quenched disk galaxies are very similar to the quenched bulges in age, metallicity and halo environment, indicating a very similar formation scenario \citep[see also][]{Zhou2020}. Future \hi\ observations for larger samples would provide more insights into their formation scenario.

\begin{figure*}
	\centering
	\includegraphics[width=0.9\textwidth]{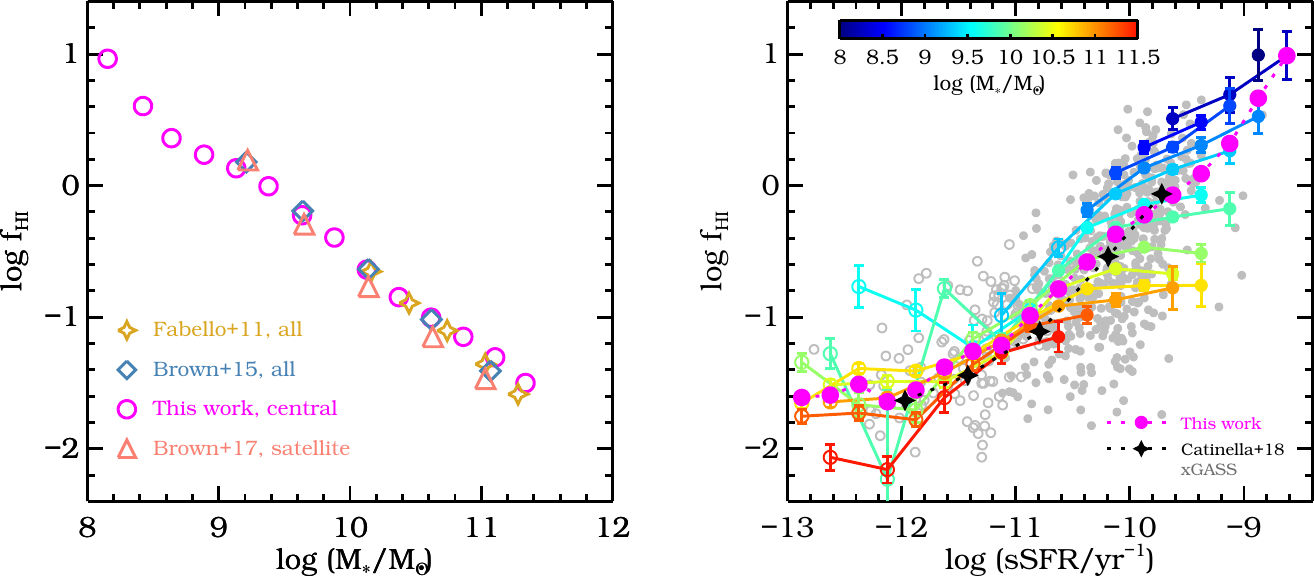}
	\caption{Comparison of the \hi\ scaling relations with literature for the $f_{\rm HI}$--$M_\ast$ (left panel) and $f_{\rm HI}$--sSFR relations (right panel). Left: our combined measurements (open circles) are compared with results from \cite{Fabello2011}, \cite{Brown2015}, and \cite{Brown2017}, using different symbols. The errorbars are omitted for clarity. Right: our measurements are shown as the circles with solid lines as in the right panel of Figure~\ref{fig:hisfrsm}, with the filled pink circles with dotted line for our averaged measurements of different stellar mass bins. The individual \hi\ detection from xGASS are shown as the filled and open gray circles for SFGs and QGs, respectively. The weighted averages of xGASS galaxies in \cite{Catinella2018} are also displayed as the black stars with dotted line.}
	\label{fig:fhissfr}
\end{figure*}

\subsection{HI Scaling Relations}\label{subsec:hiscaling}

The \hi\ scaling relations have been intensively studied in literature to understand the connection between the cold gas and other galaxy properties \citep[see e.g.,][]{Zhang2009,Catinella2010,Catinella2018,Fabello2011,Li2012,Brown2015,WangEnci2015,Wang2016}. Although we are stacking in the 2D plane of SFR--$M_\ast$, we can still compare our results with previous scaling relations of the \hi\ fraction by summing our measurements of $M_{\rm HI} (M_\ast, {\rm SFR})$ in one dimension weighted by the corresponding number of galaxies in each bin.

We show in Figure~\ref{fig:fhissfr} the comparisons of our measurements with literature for the $f_{\rm HI}$--$M_\ast$ (left panel) and $f_{\rm HI}$--sSFR relations (right panel). In the left panel, all the measurements shown are made with the \hi\ stacking technique for the ALFALFA samples at different completion stages, including the studies of \cite{Fabello2011}, \cite{Brown2015}, and \cite{Brown2017}. We note that the former two studies used all galaxies in the stacking, while the last one only stacked the satellite galaxies selected from a similar group catalog of \cite{Yang2007}. Our measurements are in very good agreement with these studies, despite the fact that we are only using the central galaxies. With the larger volume of the ALFALFA 100\% sample, we are now able to extend this relation to the lower stellar masses of $M_\ast\sim10^{8}\msun$.

This comparison shows that the average scaling relation of $f_{\rm HI}$--$M_\ast$ for central and satellite galaxies are very similar, except that the satellite galaxies have slightly smaller gas fractions at the massive end. However, as shown in \cite{Brown2017}, $f_{\rm HI}$ of the satellite galaxies strongly depends on the hosting halo mass, with satellite galaxies in massive halos suffering from severe cold gas depletion \citep[see also][]{Stevens2019}. It is different from what we find in Figure~\ref{fig:hicen} for central galaxies. The \hi\ fraction of central galaxies of $M_\ast<10^{10.5}\msun$ will increase with halo mass, while $f_{\rm HI}$ for more massive centrals show no strong dependence on the halo mass.

We compare our measurements of $f_{\rm HI}$--sSFR in the right panel of Figure~\ref{fig:fhissfr}, with the individual \hi\ detection from the xGASS sample (gray circles). The weighted averages of xGASS galaxies in \cite{Catinella2018} are also shown as the black stars with dotted line. Our measurements are shown as circles with solid lines as in the right panel of Figure~\ref{fig:hisfrsm}, and the filled pink circles with dotted line are our averaged measurements of different stellar mass bins, where we extend our measurements to a lower limit of $\log({\rm sSFR/yr^{-1}})>-13$. The overall agreement with the xGASS sample is quite good, while the individual observations have larger scatters. Thanks to the much larger sample size and the \hi\ stacking technique, we are able to reconstruct the clear trends with the stellar mass. The smaller scatters of $f_{\rm HI}$ for QGs are clear evidence of the \hi\ reduction happening in galaxies of different $M_\ast$. 

Using stacked \hi\ data of GASS galaxies, \cite{Saintonge2016} found a three-dimensional scaling relation of $\log f_{\rm HI}=0.588\log{\rm SFR}-0.902\log M_\ast+8.57$, which naturally explains the strong dependence of $f_{\rm HI}$ on $M_\ast$ for SFGs, as shown in the right panel of Figure~\ref{fig:fhissfr}. Following the same practice, we fit Equation~(\ref{eq:sfrhi}) using all our stacked data with $\log({\rm sSFR/yr^{-1}})>-12$ (59 data points weighted by the number of galaxies in each stacking), we find a scaling relation of 
\begin{equation}
	\log f_{\rm HI}=0.363\log{\rm SFR}-0.854\log M_\ast+8.159, \label{eq:fhifit_all}
\end{equation}
which would translate to ${\rm SFR}\propto M_{\rm HI}^{2.75}/M_\ast^{0.40}$, in good agreement with our estimate in Section~\ref{subsec:gsfl}. However, the scatters of this relation are significantly larger in the QGs. This is due to the fact that the slope between $f_{\rm HI}$ and sSFR is changing as galaxies quench, as clearly seen in Figure~\ref{fig:fhissfr}. As the sSFR decreases, the galaxy \hi\ content is less correlated with the star formation activity, as also seen in Figure~\ref{fig:hims}.

Therefore, to find a reasonable scaling relation between the three parameters of $f_{\rm HI}$, SFR and $M_\ast$, we fit the relations for SFGs and QGs separately and obtain,
\begin{eqnarray}
	\log f_{\rm HI,SFG}&=&0.265\log{\rm SFR}-0.794\log M_\ast+7.554\quad   \label{eq:fhifit_SFG}\\
	\log f_{\rm HI,QG}&=&0.453\log{\rm SFR}-0.624\log M_\ast+5.688\quad   \label{eq:fhifit_QG}.
\end{eqnarray}
We note that the fitting is based on the stacked measurements, not the individual measurements as in ALFALFA and xGASS. Therefore, it is not minimizing the variation of individual measurements on purpose, as done in \cite{Zhang2009} and \cite{Li2012}.  

It is noteworthy that we are still assuming a constant slope between $f_{\rm HI}$ and SFR for the SFGs, while we know the slope is changing with $M_\ast$ as shown in Figure~\ref{fig:fhissfr}. It should be regarded as an average slope for the SFGs. But the corresponding slope for the QGs does not seem to change much with $M_\ast$. For galaxies at a given $M_\ast$, it means that the global star formation law would vary from ${\rm SFR}\propto M_{\rm HI}^{3.77}$ in the star-forming population to ${\rm SFR}\propto M_{\rm HI}^{2.21}$ in the quenched population, consistent with our rough estimate of the transition slope as ${\rm SFR}\propto M_{\rm HI}^{2.7}$ in Section~\ref{subsec:gsfl}. At a given sSFR, the relation would vary from $f_{\rm HI}\propto M_\ast^{-0.53}$ to $f_{\rm HI}\propto M_\ast^{-0.17}$, i.e. the scatters with $M_\ast$ in the relation of $f_{\rm HI}$--sSFR become much smaller in the QGs. 

\begin{figure}
	\centering
	\includegraphics[width=0.45\textwidth]{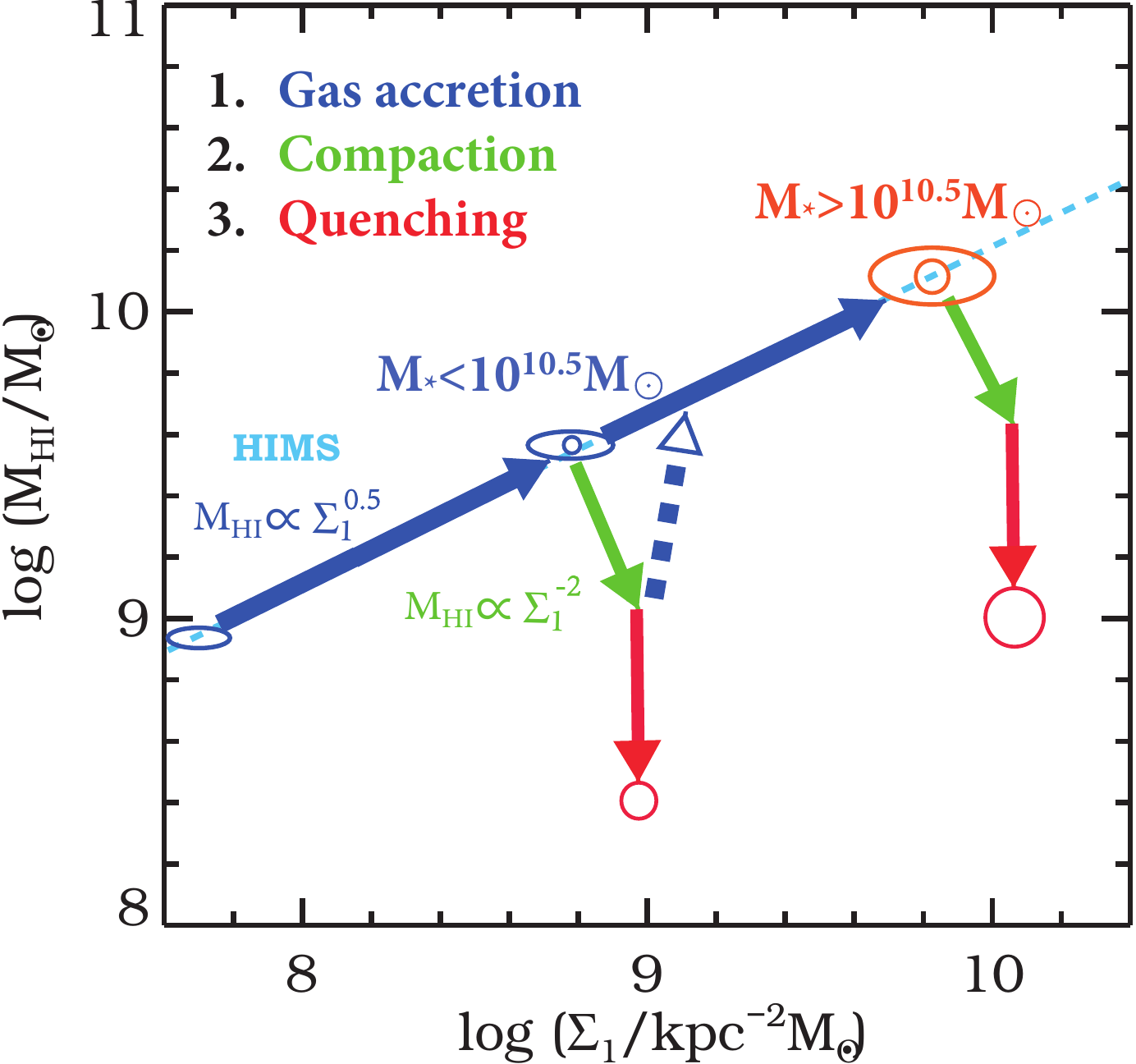}
	\caption{Illustration of the possible evolution tracks of the star-formation and quenching of central galaxies, using $M_{\rm HI}$ and $\Sigma_1$. A galaxy starting with a low mass around $10^8\msun$ will potentially undergo three phases before they fully quench, i.e. the gas accretion (blue), compaction (green) and quenching (red). The blue dotted lines represent the possible rejuvenation of the quenched low mass galaxies through new gas accretion. The results of the compaction and quenching phases are different for galaxies with masses below and above $10^{10.5}\msun$, because the environmental effect of halo quenching will take effect to shutdown the cold gas supply to the center, when their host halo masses grow above the critical value of $10^{12}\msun$. See text for details. }
	\label{fig:diagram}
\end{figure}
\section{Star formation, quenching and the relation to HI gas}\label{sec:discussion}
Our \hi\ stacking measurements establish the key role of \hi\ in sustaining star formation in low-redshift galaxies. It weakens the necessity to invoke additional mechanisms that suppress or prevent the \hi\ from streaming in, or from cooling and forming molecular gas and stars. As we show in Sections~\ref{subsec:gsfl} and~\ref{subsec:hiscaling}, the star formation and quenching of these low-redshift central galaxies are generally governed by the global star formation law of ${\rm SFR}(M_{\rm HI},M_{\ast})$ as in Equation~(\ref{eq:sfrhi}). We note that the scaling coefficients in this equation only vary slightly with the star formation status. With an average scaling relation of ${\rm SFR}\propto M_{\rm HI}^{2.75}/M_\ast^{0.40}$, it emphasizes the important role of \hi\ reservoir in regulating the star formation. 

These results are consistent with recent works based on the ALMaQUEST survey quantifying the role of molecular gas in regulating the star formation at low redshifts with resolved observations \citep{Lin2019,Ellison2020,Ellison2020a,Ellison2021,Ellison2021a}, as well as the  xCOLD-GASS survey using the global molecular measurements \citep{Saintonge2016}. They found that both the resolved and integral SFMS can be fully explained by a combination of the molecular abundance and the molecular star formation law. As \hi\ needs to be converted to \hj\ before forming stars, these results on the molecular gas perfectly fits and fills the physical gap between \hi\ and SFR in our relations. The similar behaviors of our global SFR-\hi\ scaling relation to the resolved ones \citep{Bigiel2008,Bigiel2010}, and to the SFR-molecular gas scaling relations, strongly infer the quasi-equilibrium baryonic flow where \hi\ is smoothly converted to \hj\ and then to stars \citep[as also suggested in][]{Saintonge2016} for the majority of star-forming galaxies at low redshifts. It suggests that the star formation and quenching is regulated by the availability of the \hi\ reservoir, rather than the stellar or halo masses.

However, the detailed physics involved in changing the \hi\ reservoir is hard to clearly identify in observation. It may include \hi\ accretion, outflow, cooling of hot gas, compaction, mergers, stellar and AGN feedback, as suggested in literature \citep[see][for reviews]{Peroux2020,Tacconi2020}. From \hi\ measurements of the bulge- and disk-dominated galaxies, we do not find any strong evidence of morphological quenching, as both populations follow the same global star formation law as shown in Figure~\ref{fig:hisfrmorph}. There is no significant difference between the \hi\ reservoir in bulge- and disk-dominated SFGs (Figure~\ref{fig:hisfrmorph}), as also found by \cite{Cook2019} and \cite{Cook2020}. The massive quenched disks show relatively higher \hi\ masses than those of the quenched bulges, due to their slightly higher SFRs. 

The scaling relations of $M_{\rm HI}$ with $\Sigma_1$ and $M_{\rm h}$ seem to be consistent with the compaction-triggered quenching scenario presented in earlier studies \citep[see e.g.,][]{Dekel2014,Zolotov2015}. But the increasing of $\Sigma_1$ is not equivalent to gradual dominance of the bulge. Due to the large scatters between $\Sigma_1$ and $B/T$, many disk galaxies can still have large values of $\Sigma_1$. As shown in Figure~\ref{fig:hisfrmorph}, the correlation between $M_{\rm HI}$ and $\Sigma_1$ persists in the disk galaxies, with a very similar slope as well as intercept as those of the bulge-dominated systems. The consistent slopes between $M_{\rm HI}$ and $\Sigma_1$ in the transition from SFMS to the quiescent phase at all stellar masses (Figure~\ref{fig:hisigma2}), further confirms that their correlation is not simply caused by the bulge growth.

However, we emphasize that the strong correlation between $M_{\rm HI}$ and $\Sigma_1$ does not necessarily indicate the causal relation of compaction and quenching, because other physical mechanisms may possibly cause the quenching, as well as the compaction. We only treat the compaction-triggered quenching as a potential candidate to summarize the scaling relations presented in this paper. The overall possible evolution picture of central galaxies can be reasonably described using the parameters of $M_{\rm HI}$, $\Sigma_1$ and $M_{\rm h}$, as illustrated in Figure~\ref{fig:diagram}. The galaxies starting from a low mass of around $10^8\msun$ will potentially undergo three phases before they finally quench, unified in the scenario as follows.

\begin{enumerate}
	
	\item Cold gas accretion
	
	As we see in Figure~\ref{fig:hicen}, galaxies with $M_\ast\sim10^{8}\msun$ will increase their \hi\ reservoir smoothly with increasing halo mass, in a rate around $M_{\rm HI}\propto M_{\rm h}$. Their \hi\ mass is less correlated with the stellar mass, when the gas inflow timescale becomes smaller than the star formation timescale \citep{Tacchella2016}. The dependence of $M_{\rm HI}$ on $\Sigma_1$ in this stage is mainly driven by the stellar mass growth, with the diffuse star-forming cores gradually becoming dense \citep{Barro2017}.
	
	The uniformity of Equation~(\ref{eq:sfrhi}) throughout the sample space suggests that such an inflow almost always happens, possibly with a flow rate proportional to the abundance of \hi. A positive correlation of $\Delta\log{\rm SFR}$ with the mass and averaged surface density of \hi\ in the inner optical disks is found by \cite{Wang2020} using disk-dominated galaxies of xGASS. 
	
	\item Compaction and post-compaction
	
	The increase of the gas surface density, as well as the slow bulge growth, will make the Toomre instability parameter Q quickly decrease, leading to the disk compaction. The compaction of gas naturally results in the significant increase of star formation because SFR is super-linearly correlated with \hi\ mass density in the star-forming regions. The galaxy central core is built up, as can be seen from the slightly positive slope of the $\Delta\Sigma_1$--$\Delta\log M_{\rm HI}$ relation when $\Delta\log M_{\rm HI}>0$ (Figure~\ref{fig:hisigma1}). We do not measure the \hi\ surface densities here, but \cite{Wang2020} showed that for at least the disk-dominated galaxies, the averaged \hi\ surface density increases with $\Delta\Sigma_1$ at a given $\Delta\log{\rm SFR}$. 
	
	The end of compaction leads to the decrease of \hi\ mass due to the quick depletion from star formation, outflow and feedback mechanisms. The galaxy thus moves to the phase of post-compaction, and star formation gets suppressed to a local minimal level before the next round of compaction (unless the current round results in permanent quenching), as determined by the star formation law. The slope of $d M_{\rm HI}/d\Sigma_1$ becomes much steeper in this phase ($M_{\rm HI}\propto\Sigma_1^{-2}$ when $\Delta\log M_{\rm HI}<0$, Figure~\ref{fig:hisigma1}). We note that, such a relation does not necessarily link to a continuous evolutionary track along this slope. As suggested in \cite{Tacchella2016}, low-mass galaxies may experience a couple of gas replenishment, compaction and depletion cycles before they finally quench. Galaxies going through different phases in their evolution cycles may mix up together in this area to form the relation of $M_{\rm HI}\propto\Sigma_1^{-2}$. That is, the position of those post-compaction galaxies may mix with newly replenished galaxies before the next round of compaction, which would smooth the $\Delta\Sigma_1$--$\Delta\log M_{\rm HI}$ relation. 
	
	The compaction result is slightly different for galaxies reaching the critical mass of $10^{10.5}\msun$. These galaxies live in halos of masses larger than around $10^{12}\msun$, where the virial shocking heating and AGN feedback will take effect to shut down the cold gas supply to the halo center \citep{Dekel2006,Dekel2014}, as seen in Figure~\ref{fig:hicen}. 
	
	\item Quenching
	
	Once the galaxy center reaches the maximum gas compactness, it will start to quench. The slope between SFR and $M_{\rm HI}$ will gradually change during the compaction and quenching phase (shown in Figure~\ref{fig:hisfrsm}), with an average slope around ${\rm SFR}\propto M_{\rm HI}^3$. Thus, the SFR will be significantly decreased even if $M_{\rm HI}$ is only lowered by around 0.6~dex. As shown in Figure~\ref{fig:hicen}, galaxies below $10^{10.5}\msun$ might experience rejuvenation due to the growth of \hi\ reservoir with the halo mass (represented by the dotted blue lines in Figure~\ref{fig:diagram}), when the gas replenishment timescale is smaller than the depletion timescale \citep[see also][]{Tacchella2016}.
\end{enumerate}

The above scenario has been presented with more sophisticated details in the previous theories and simulations  \citep[e.g.,][]{Dekel2006,Dekel2014,Tacchella2016}. But here for the first time, we put the three steps coherently together based on one observational dataset, uniquely linking the $M_{\rm HI}$--$M_{\rm h}$ relation (which sets the \hi\ abundance in the accretion and quenching phases) with the $M_{\rm HI}$--SFR--$M_\ast$ scaling relation (which determines the rate of \hi\ converting into stars at a given abundance in the compaction and post-compaction phase). It is interesting that the combination of physics on the very large (halo) and very small (star-forming clouds) scales basically determines the integral star-forming status of a galaxy.

\section{Conclusions}\label{sec:summary}
With accurate stacked measurements of the \hi\ signals in the overlap regions between ALFALFA and SDSS DR7, we make it possible to quantitatively investigate the dependence of \hi\ mass on halo mass, stellar mass, SFR, and central stellar surface density, for both the star-forming and quenched central galaxy populations. 

Our main conclusions are summarized as follows.

\begin{itemize}
	\item The shapes in the $M_{\rm HI}$--$M_{\rm h}$ and $M_{\rm HI}$--$M_\ast$ relations are remarkably similar for SFGs and QGs, with the quenched galaxies having consistently lower \hi\ masses around 0.6--1~dex. It indicates that neither the stellar mass nor the halo mass is the direct cause of quenching. The \hi\ masses of galaxies with $M_\ast<10^{10.5}\msun$ strongly increase with $M_{\rm h}$, while for more massive galaxies the \hi\ reservoir does not vary with $M_{\rm h}$. The SFGs form a tight HIMS, with $M_{\rm HI}\propto M_\ast^{0.42}$, while the QGs show similar dependence with slightly larger scatters.
	\item Central galaxy star formation and quenching are generally regulated by the \hi\ reservoir, following an average global star formation law of ${\rm SFR}\propto M_{\rm HI}^{2.75}/M_\ast^{0.40}$. The correlation would be better fitted when separating the relations for SFGs and QGs as in Equations~(\ref{eq:fhifit_SFG}) and (\ref{eq:fhifit_QG}), due to the slight slope changes in the two populations. The observed gas fraction $f_{\rm HI}$ shows large scatters with sSFR for the SFGs, mainly caused by the stellar mass dependence.    
	\item The \hi\ masses of central galaxies are decreased once they drop off the SFMS, confirming that the cold gas depletion is the main culprit of quenching. There is also a strong and consistent correlation of $M_{\rm HI}\propto\Sigma_1^{-2}$ for quenching galaxies in this phase, supporting the idea of compaction-triggered quenching. 
	\item There is no significant differences in the \hi\ reservoir for the bulge- and disk-dominated galaxies in both the star-forming and quenched populations. Galaxies with different morphology types both follow the same global star formation law, corroborates that depletion of the \hi\ reservoir is the sufficient and necessary requirement for quenching, regardless of morphology.
	\item Our results are consistent with the compaction-triggered quenching scenario, suggesting that the evolution of central galaxies generally goes through three different phases of cold gas accretion, compaction and quenching. 
	
\end{itemize}

\acknowledgments
This work is supported by the National Key R\&D Program of China (grant No. 2018YFA0404503), National SKA Program of China (No. 2020SKA0110100), National Science Foundation of China (Nos. 11922305, 11773049, 12073002, 11721303), and Spanish Science Ministry ``Centro de Excelencia Severo Ochoa'' program under grant SEV-2017-0709. MGJ was supported by a Juan de la Cierva formaci\'{o}n fellowship (FJCI-2016-29685). MGJ also acknowledge support from the grants AYA2015-65973-C3-1-R and RTI2018-096228-B-C31 (MINECO/FEDER, UE).

We thank the anonymous reviewer for the helpful comments that significantly improve the presentation of this paper. We acknowledge the work of the entire ALFALFA team for observing, flagging and performing signal extraction. We thank Xiaoyang Xia and Cheng Li for helpful discussions. We acknowledge the use of the High Performance Computing Resource in the Core Facility for Advanced Research Computing at the Shanghai Astronomical Observatory.

Funding for the SDSS and SDSS-II has been provided by the Alfred P. Sloan Foundation, the Participating Institutions, the National Science Foundation, the U.S. Department of Energy, the National Aeronautics and Space Administration, the Japanese Monbukagakusho, the Max Planck Society, and the Higher Education Funding Council for England. The SDSS Web Site is http://www.sdss.org/.

\facility{Arecibo, Sloan}
\bibliographystyle{aasjournal}

\end{document}